\documentclass[aps,prl,reprint,superscriptaddress]{revtex4-1}
\usepackage{amsmath}
\usepackage{graphicx}
\usepackage{color}
\usepackage{soul}
\usepackage{siunitx}
\usepackage{braket}

\def \Kp {\textrm{K}^\prime}

\begin{document}

\author{Sudipta Kundu}
\author{Mit H. Naik}
\altaffiliation[Present address: ]{Department of Physics, University of California at Berkeley, California 94720, USA and Materials Sciences Division, Lawrence Berkeley National Laboratory, Berkeley, California 94720, USA.}
\author{H. R. Krishnamurthy}
\author{Manish Jain}
\affiliation{Center for Condensed Matter Theory, Department of Physics, Indian Institute of Science, Bangalore 560012, India}

\title
{Moir\'{e} induced topology and flat bands in twisted bilayer WSe$_2$: A first-principles study}

\begin{abstract}
We study the influence of strong spin-orbit interaction on the formation of
flat bands in relaxed twisted bilayer WSe$_2$.  Flat bands, well separated in
energy, emerge at the band edges for twist angles ($\theta$) near
0\si{\degree} and 60\si{\degree}.  For $\theta$ near 0\si{\degree}, the
interlayer hybridization together with a moir\'{e} potential determines the
electronic structure.  The bands near the valence band edge have nontrivial
topology, with Chern numbers equal to +1 or $-$1.  We propose that the nontrivial topology of the first band can be
probed experimentally for twist angles less than a critical angle of
3.5\si{\degree}.  For $\theta$ near 60\si{\degree}, the flattening of the bands
arising from the K point of the unit cell Brillouin zone is a result of atomic
rearrangements in the individual layers.  Our findings on the flat bands and
the localization of their wavefunctions for both ranges of $\theta$ match well
with recent experimental observations.
\end{abstract}

\maketitle

Twisted bilayer transition metal dichalcogenides (TMD) have recently gained
attention as potential platforms for hosting correlated phases and novel
excitonic properties
\cite{wse2_expt1,wse2_expt2,wse2_expt4,wse2_expt5,wse2_expt6,wse2_expt7,tmd1,tmd3,tmd6,tmd7,tmd8,tmd9,lado,hetero1,hetero3,hetero4,hetero5,hetero7,hetero8,hetero9,hetero10}.
The absence of a 'magic angle' in twisted TMDs
\cite{naik1,naik2,macdonald_mos2,rubio_mos2,yang_mos2} makes the experimental realization
of flat bands \cite{wse2_expt1,wse2_expt2} easier compared to twisted bilayer
graphene. Large spin-orbit coupling can strongly influence the electronic
structure of the moir\'{e} superlattice (MSL), and cause the bands to have
nontrivial topological character.  Within the TMD family, WSe$_2$ is a
prototypical example with large spin-orbit coupling, making twisted WSe$_2$
bilayers (tWSe$_2$) especially interesting.

Recent spectroscopic imaging and transport measurements in tWSe$_2$ show
evidence of flat bands \cite{wse2_expt1,wse2_expt2,wse2_expt3}. As we discuss in detail in this letter, the flat bands
at the valence band (VB) edge in the MSL of tWSe$_2$ arise from the bands near
the K points of the unit cell Brillouin zone (UBZ), {\em unlike} in other TMDs like MoS$_2$.
The origin and electronic structure of the flat bands arising from the K point are different from
those due to the $\Gamma$ point; the latter are primarily determined
by the interlayer hybridization, whereas the former are a result of the ``moir\'{e}
potential'', the additional effective potential generated due to the relaxation of the atoms upon twisting.
The spin character of the flat bands is determined by the spin-valley locking in WSe$_{2}$.
Due to the large spin-orbit splitting and the spin-valley locking at K/$\Kp$-valley, tWSe$_2$
is a good platform  for exploring opportunities for spintronics as well as valleytronics.

Most theoretical studies on tWSe$_2$ to date 
\cite{wse2_thry1,wse2_thry2,wse2_thry3,wse2_thry4,wse2_thry5,wse2_lfu} have been based
on continuum models \cite{continm,kaxiras}.  In this letter, we study flat
bands in {\em relaxed} MSL of tWSe$_2$ using density functional theory (DFT).
We focus our study on twist angles $\theta$ near 0\si{\degree}, ranging from
7.3\si{\degree} to 1.89\si{\degree}, and near 60\si{\degree}, ranging from
52.7\si{\degree} to 58.11\si{\degree}. We find several well-separated flat
bands near the band edges in tWSe$_2$ for both ranges of $\theta$.
The spin-orbit splitting of the monolayer bands is preserved
in tWSe$_2$. {\em Both} the moir\'{e} potential and the interlayer hybridization
govern the electronic structure of tWSe$_2$ with $\theta$ near 0\si{\degree}; in contrast,
the flat bands in tWSe$_2$ with $\theta$ near 60\si{\degree} primarily result from the moir\'{e}
potential alone. Flat bands near the VB edge for $\theta$ near 0\si{\degree}
are topologically nontrivial \cite{continm}. This stems from the nonzero Berry curvature
at the K-valleys of the UBZ. We fit the bands to the continuum model \cite{continm}; the obtained parameters
thus include the effect of structural relaxation of the MSL. We also find flat bands arising from the
$\Gamma$ point of the UBZ {\em inside the VB continuum} for both ranges of $\theta$. The
spatial localization of these flat bands is in excellent agreement with recent
scanning tunnelling measurements (STM) \cite{wse2_expt1}. 

We note that larger deviations of $\theta$ from 0\si{\degree} or 60\si{\degree}
lead to smaller moir\'{e} supercells. In such
systems, the effect  of kinetic energy is more prominent compared to the moir\'{e} effect,
and the bands in the moir\'{e} Brillouin zone (MBZ) are primarily a result of the band folding.
Conversely, the closer $\theta$ is to 0\si{\degree} or 60\si{\degree}, the larger the moir\'{e}
supercell, and the flatter the bands, increasing the possibilities for interesting strong
correlation phenomena, with scope for additional richness in the presence of bands with
nontrivial topology. Hence, we focus in this letter on tWSe$_2$ with $\theta$ near 0\si{\degree}
and 60\si{\degree}.  It is to be noted that 30\si{\degree} tWSe$_2$ can give rise to
quasicrystalline order \cite{stampfli,tblg30}, and we hope to study such
systems in the future.

Our electronic structure calculations were performed using the atomic-orbital basis
as implemented in the SIESTA package \cite{siesta}.  The commensurate tWSe$_2$
structures are generated using the Twister code \cite{naik1} and relaxed using
classical force fields as implemented in LAMMPS \cite{lammps,sw,kc,kc_naik} (for
other details, see Supplemental material (SM) \cite{sm_ref}).  The relaxation
patterns agree well with previous studies on other TMDs
\cite{naik2,falko_cont,struc,kaxiras_cont,struc}.  MSLs with $\theta$ near
0\si{\degree} consist of regions with three high symmetry stackings: AA,
AB, and BA (also known as 3R, B$^{\textrm{W/Se}}$, and B$^{\textrm{Se/W}}$ respectively), as
well as bridge (Br) regions (the transition regions from one such stacking to
another) (see SM \cite{sm_ref}).  MSLs with $\theta$ near 60\si{\degree}
consist of three other high symmetry stackings: AA$^{\prime}$, A$^{\prime}$B and AB$^{\prime}$ (also known as 2H, B$^{\textrm{Se/Se}}$, and
B$^{\textrm{W/W}}$ respectively), as well as the bridge (Br) regions.  We have 
performed separate calculations of WSe$_2$ bilayers with the above high symmetry stackings, using 
DFT \cite{qe,pz,kpt,dfc09,pbe}as well as GW techniques \cite{sm_ref,BGW,gpp,cellslab,deslippe} and find the VB maximum to be
at the K point of the UBZ.
It is to be noted here that the the relative
positions of the VB edge at $\Gamma$ and K and hence the position of VB maximum in bilayer WSe$_2$
depend strongly on the interlayer separation \cite{sm_ref,k-g1,k-g2,layerdep}.
Recently, there has been a study
which suggests that the flat bands in tWSe$_2$ at VB edge are due to $\Gamma$
point of UBZ \cite{johannes}. Nevertheless, the experimental findings on
tWSe$_2$ show that the bands at the VB edge arise from K point of the UBZ, and
we find the same.

\begin{figure}
\centering
\includegraphics[scale=0.27]{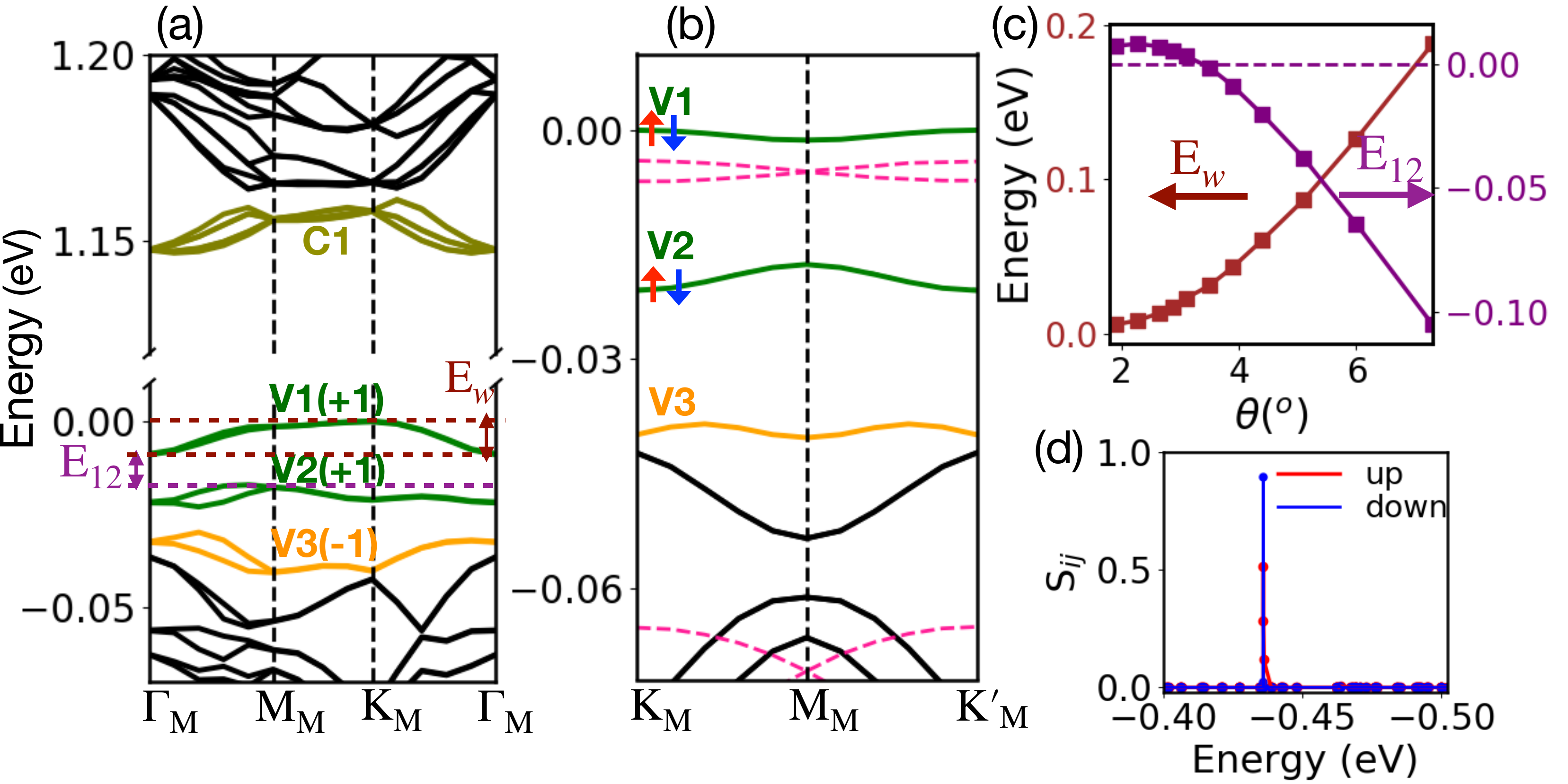}
\label{fig1}
\caption{(a): Band structure of 2.28\si{\degree} tWSe$_2$ along the path $\Gamma_{\textrm{M}}-\textrm{M}_{\textrm{M}}-\textrm{K}_{\textrm{M}}- \Gamma_{\textrm{M}}$ with the VB maximum set to zero.
(b): Bands near the VB edge along the path K$_{\textrm{M}}-\textrm{M}_{\textrm{M}}-\Kp_{\textrm{M}}$.
The spin orientations of the first two bands, each doubly degenerate, are indicated by red and blue arrows.
The pink dashed lines represent the band structure of the relaxed-MSL monolayer. 
For ease of viewing, the VB maximum of the monolayer has been set to -4 meV.
(c): Variation of E$_W$ (brown line) and E$_{12}$ (purple line) with $\theta$.
(d): Variation of S$_{ij}$ with energy. The red and blue lines represent spin-up and spin-down, respectively.}
\end{figure}

\begin{figure}
\centering
\includegraphics[scale=0.45]{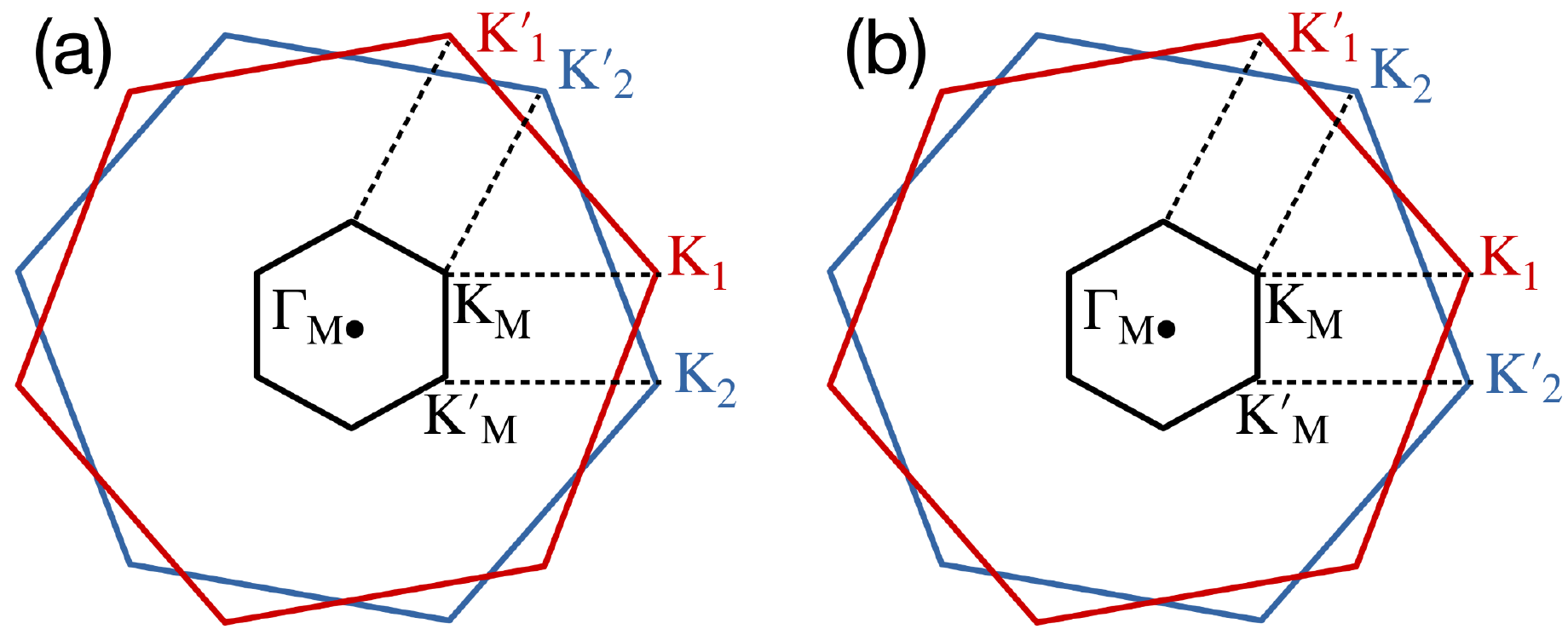}
\caption{(a): The schematic for the folding relation of the UBZ to the MBZ
for tWSe$_2$ with $\theta$ near 0$^{\circ}$. The small black hexagon represents
the MBZ, and the red and blue hexagons represent the UBZ for the two layers, respectively. The K and
$\Kp$ points in all the Brillouin zones are shown. The subscripts 1 and 2 stand for the
two layers. (b): Same as above for tWSe$_2$ with $\theta$ near 60$^{\circ}$.}
\end{figure}

{\bf{\textit{Electronic structure of tWSe$_2$ with $\theta$ near 0\si{\degree}:}}}
Fig. 1 presents our results for the band structures of tWSe$_2$ for $\theta$
near 0\si{\degree}. For such $\theta$ values, bands near the K$_{\textrm{M}}$
point of the moir\'{e} MBZ (subscript M denotes
\textbf{k}-points in the MBZ) arise from the monolayer bands near the K and $\Kp$
points of the UBZ of the two layers. The zone folding relation is
illustrated in Fig. 2(a). As the monolayer VB maximum at the K and $\Kp$
points have antiparallel spins, the VB maximum at the K$_{\textrm{M}}$ point is
doubly degenerate, with antiparallel spins.  Fig. 1(a) shows the bands for the
2.28\si{\degree} tWSe$_{2}$.  Two-fold degenerate bands with opposite spins are
evident near the VB edge (V1, V2, V3). These first few VBs are shown along the
path $\textrm{K}_{\textrm{M}}-\textrm{M}_{\textrm{M}}-\Kp_{\textrm{M}}$
in Fig. 1(b). There is always a gap between V1 and V2, even at the
$\textrm{M}_{\textrm{M}}$ point, due to the interlayer hybridization (Fig.
1(b)); this is clear from the fact that the bands for the relaxed-MSL monolayer
(i.e., without interlayer interaction but with the structural relaxation effects of
the MSL included) along the same path (pink dashed line
in Fig. 1(b)) show a band crossing at $\textrm{M}_{\textrm{M}}$.

To study the topological aspects of the band structure, we have calculated the
Chern number (C$_n$) for the first few bands near the VB edge.
As the bands are doubly degenerate with antiparallel spin,
we differentiate between the flat bands by following the spin. For this purpose we calculate the
expectation value of the Pauli matrices for the first few bands at the VB edge and
find that $\langle\sigma_z\rangle$ is $\sim \pm$ 1, and $\langle\sigma_x\rangle$
and $\langle\sigma_y\rangle$ are nearly zero. We consider the band with $\langle\sigma_z\rangle$ = +1 ($-1$)
as the spin-up (spin-down) band. We generate the Bloch wavefunctions
using DFT on a regular k-grid in the MBZ
and compute the C$_n$'s \cite{cherndft1,cherndft2,sm_ref}.
The larger $\theta$ is, the denser is the k-grid we have used. For $\theta=$1.89\si{\degree},
we compute the $C_n$ using a 6$\times$6 k-grid in the MBZ and the
$C_n$ are $+$1, $+$1, $+$1 respectively, for the V1, V2 and V3 spin-up bands.
For 2.28\si{\degree} $\geq \theta >$ 4.4\si{\degree},
$C_n$ are $+$1, $+$1, $-$1 for V1, V2 and V3 respectively.
For $\theta$ $\geq$ 4.4\si{\degree} we
report C$_n$ for the first two isolated bands ($+1$ for both V1 and V2) only. The spin-down bands
have C$_n$'s with opposite signs. It is important to note that the first band is topologically trivial when calculated for
unrelaxed tWSe$_2$, and the 2nd band overlaps with other bands.
The $\theta$ variation of the bandwidth of V1 (E$_{W}$) and the
minimum energy gap between V1 and V2 (E$_{12}$) are shown in Fig.  1(c). While E$_W$
increases monotonically with $\theta$, E$_{12}$ becomes negative at
3.5\si{\degree}. While nonzero C$_n$ is found for a large range of $\theta$,
the negative E$_{12}$ limits the range of angles for observing quantum
spin-Hall insulating states of the first band in tWSe$_2$ \cite{hall1,hall2}. Furthermore,
the quantum spin-Hall state can be probed for the first two bands till 4.4\si{\degree}. We have compared our DFT
results to the continuum model \cite{continm}. While the latter gives the Chern
number of the first two bands consistent with DFT
for all $\theta$, it is unable
to describe the DFT-calculated C$_n$ of the third band \cite{sm_ref}. 

Fig. 1(a) also shows six split-off bands (C1) near the conduction band (CB)
edge.  These bands are degenerate at $\Gamma_{\textrm{M}}$ with three having spin-up
and other three spin-down.  The Q valley of the UBZ gives rise to these bands near
the CB minima in the MBZ. The bands near the CB minima at alternate Q
points of the UBZ have antiparallel spins.  This determines the spin-character
of C1 as well as of the next set of bands near the CB minimum in the MBZ, which are
also 6-fold degenerate at $\Gamma_{\textrm{M}}$.

In order to clarify the nature of the spin-orbit splitting in the MSL, we have calculated the matrix
elements of the spin-raising/lowering operator $S^{\pm} (= S_{x} \pm iS_{y})$:
S$_{ij}$ = $\langle\psi_{i}^{\textrm{K}_{\textrm{M}}}\lvert
S^{\pm}\lvert\psi_{j}^{\textrm{K}_{\textrm{M}}}\rangle$ where
$\psi_{i}^{\textrm{K}_{\textrm{M}}}$, $\psi_{j}^{\textrm{K}_{\textrm{M}}}$ are
two wavefunctions at the $\textrm{K}_{\textrm{M}}$ point of MBZ.
We calculate the $S_{ij}$ for $i$  corresponding to the two highest energy states of VB edge 
(at the $\textrm{K}_{\textrm{M}}$ point) and $j$ to the
states within the energy range 0--500 meV below.
We find states
436 meV below the VB maximum for which S$_{ij}$ is large (Fig. 1(d)). This implies
that the atomic spin-orbit splitting remains essentially unchanged in MSLs.

\begin{figure}
\centering
\includegraphics[scale=0.32]{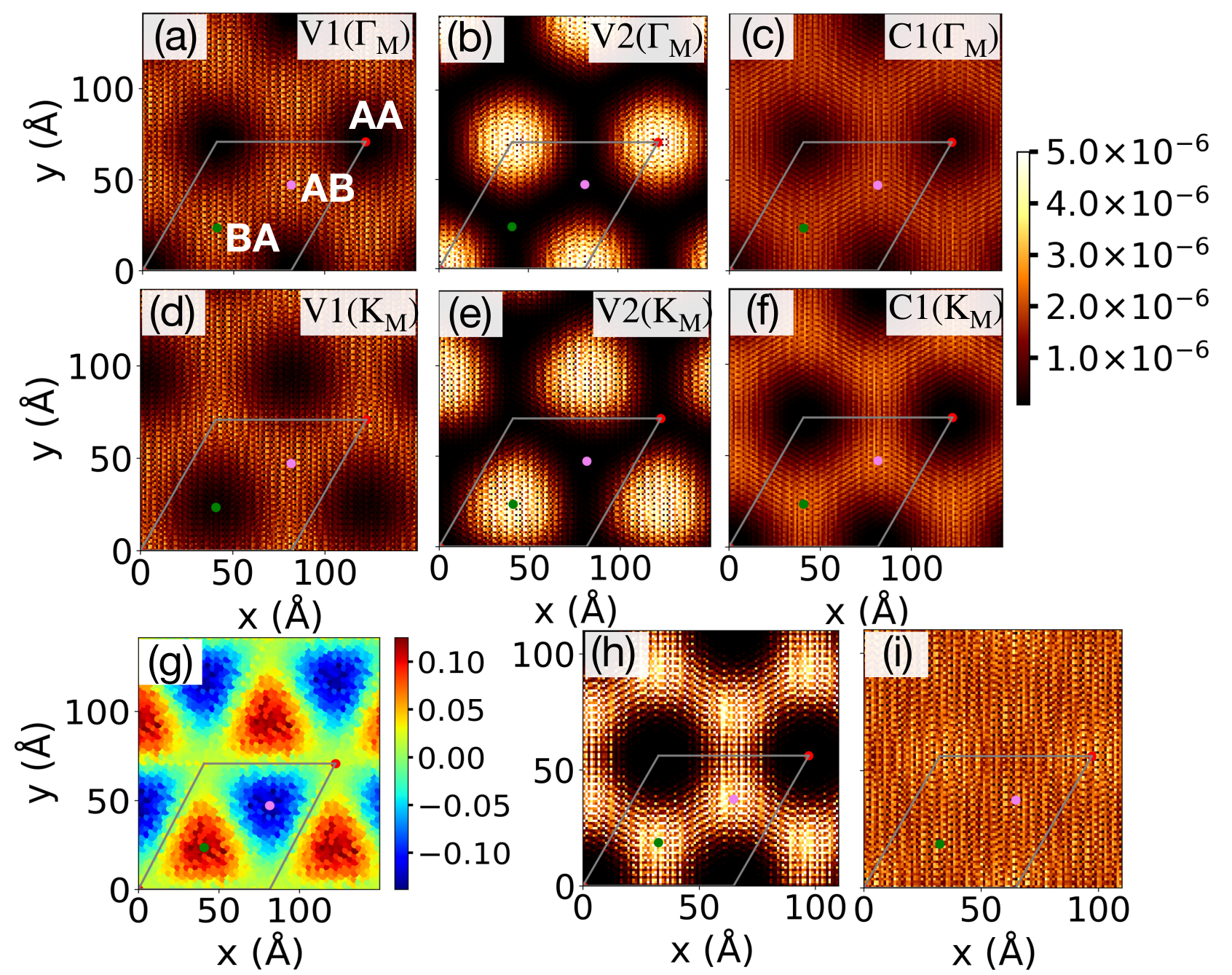}
\label{fig2}
\caption{(a), (b), and (c) ((d), (e), and (f)): Distribution of  $|\psi_{\Gamma_{\textrm{M}}}(\textbf{r})|^2$ ($|\psi_{\textrm{K}_{\textrm{M}}}(\textbf{r})|^2$) averaged
along the out-of-plane ($z$) direction for the
V1, V2 and C1
bands for $\theta=$2.28\si{\degree} tWSe$_2$.
(g): Moir\'{e} potential (in eV) of the relaxed-MSL monolayer.
(h): LDOS (normalized) of the first flat band arising from the $\Gamma$ point of the UBZ.
(i): LDOS of the V1 states arising from the K point of the UBZ. The high symmetry stacking regions are as marked in panel (a).}
\end{figure}

\textit{Wavefunction localization:}
Next, we discuss the localization of the wavefunctions of the bands mentioned
above, using false color plots of the probability $|\psi_{\textbf{k}}(\textbf{r})|^2$ 
inside the MSL, for special choices of \textbf{k}.
(The localization characteristics vary with the choice of \textbf{k}.)  Figs. 3(a) and (b)
depict the localization of the wavefunctions corresponding to V1 and V2
respectively at the $\Gamma_{\textrm{M}}$ point; the V1 wavefunction shows a hexagonal pattern avoiding
AA stacking regions and occupying the AB and BA regions, while the V2 wavefunction is
localized in the AA regions.  Figs. 3(d) and (e) show the localization of
V1 and V2 at K$_{\textrm{M}}$ for {\em one spin orientation} (say spin-down);
V1 localizes on AA and AB regions, while V2 does so on BA regions.  At K$_{\textrm{M}}$, the 
localization of the other spin (spin-up, not shown) is complementary; V1
localizes on AA and BA regions and V2 on AB regions.  Figs. 3(c)
and (f) depict the localization of the C1 wavefunctions at $\Gamma_{\textrm{M}}$ and
K$_{\textrm{M}}$ respectively. In both cases C1 localizes on AB and BA regions as the 
CB minimum at the Q valley in the UBZ has the lowest energy for AB and BA stackings.

A complimentary picture of the localization is provided by the coarse-grained
``moir\'{e} potential'' ($V_{M}$) \cite{naik2} of a relaxed MSL monolayer.  The
coarse-grained potential is obtained by averaging the potential in a Voronoi
cell centered at the W atoms ($x_{W},y_{W}$) in each layer:
\begin{equation}
V_{M}(x_{W},y_{W}) = \frac{\int_{\Omega_{Vor}(x_{W},y_{W})}V(x,y,z)d\textbf{r}}{\Omega_{Vor}(x_{W},y_{W})} - \bar{V}.
\end{equation}
Here $V(x,y,z)$ is the local potential obtained from the DFT calculation and
$\bar{V}$ is its mean.  Fig. 3(g) depicts the moir\'{e} potential of one of the layers.
The variation of the potential is driven by the strain and relaxation patterns.
The resulting V$_M$ shows an alternate arrangement of maxima and minima in a
hexagonal shape centred on the AA regions. The other layer has a similar
pattern with the positions of maxima and minima interchanged.  The moir\'{e}
potential along with the interlayer hybridization determines the electronic
structure of the relaxed MSL of tWSe$_2$ for $\theta$ near 0\si{\degree}.

\textit{Comparison with experiment:}
STM images of 3\si{\degree} tWSe$_2$ \cite{wse2_expt1} showed signatures of flat bands derived from the
$\Gamma$ point of the UBZ and the local density of states (LDOS) corresponding to
these bands .  We find flat bands arising from the $\Gamma$
point of UBZ {\em inside the VB continuum} \cite{sm_ref}.  The LDOS associated
with these $\Gamma$-derived states localizes strongly on AB, BA and Br regions,
forming a hexagonal pattern (Fig.  3(h)). This is in excellent agreement with
the STM image of the $\Gamma$-derived flat band \cite{wse2_expt1}. In contrast, the LDOS of
V1, which arises due to the K point of the UBZ, and is shown in Fig.  3(i),
is delocalized in the moir\'{e} unit cell, also consistent with the
experimental findings \cite{wse2_expt1}.

\begin{figure}[h]
\centering
\includegraphics[scale=0.30]{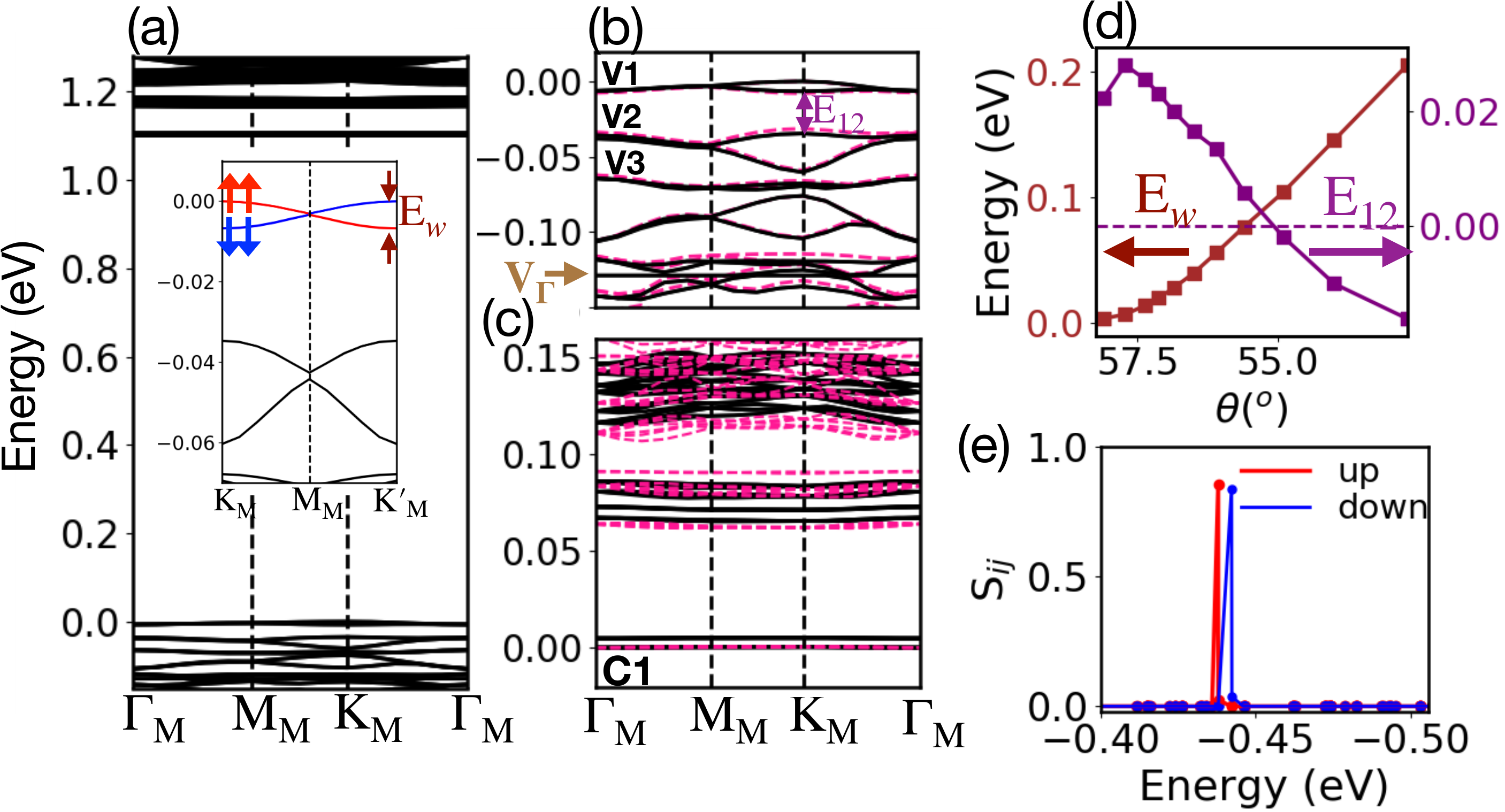}
\label{fig3}
\caption{(a): Band structure of 57.72\si{\degree} tWSe$_2$. The VB maximum is set to zero. Inset: Bands near the VB edge along the
$\textrm{K}_{\textrm{M}}-\textrm{M}_{\textrm{M}}-\Kp_{\textrm{M}}$ path. The spin-characteristic of the first two bands
are shown in red and blue colours. (b) and (c): Bands near the VB and CB edges in (a) for the relaxed-MSL bilayer (black)
and the relaxed-MSL monolayer (pink dashed). The VB maxima and CB minima are set to zero as appropriate.
(d): E$_W$ and E$_{12}$ vs the twist angle  $\theta$.
(e): S$_{ij}$ vs energy for spin-up (red) and spin-down (blue) states.}
\end{figure}

{\bf {\textit{Electronic structure of tWSe$_2$ with $\theta$ near 60\si{\degree}:}}}
We next discuss our findings for the electronic structure and the
wavefunctions of the relaxed MSLs obtained for $\theta$ near
60\si{\degree}. 

Fig. 4(a) depicts the band structure for the 57.72\si{\degree}
MSL, with the bands near the VB and CB edges being shown more clearly in Fig.s 4(b) and 4(c). 
Several well-separated flat bands emerge at the VB edge (Fig. 4(b)). These
bands are four-fold degenerate at the $\Gamma_{\textrm{M}}$ point of the MBZ,
and split into two sets of two-fold degenerate bands at the
$\textrm{K}_{\textrm{M}}$ point, with the spins of each set being parallel (inset of
Fig. 4(a)) but antiparallel to the spins of the other set.  This is because the first set
of bands at the $\textrm{K}_{\textrm{M}}$ point of the MBZ arise from the VB maxima at the K
points of the UBZ of the two layers, which
have the same spin.  The second set of bands with the opposite spin at
the $\textrm{K}_{\textrm{M}}$ point arises from the VBs at the $\Kp$ (modulo a reciprocal
lattice vector of the MSL) point of the UBZ of the two layers. This
folding relation is shown in Fig. 2(b). The bands at the VB edge for
$\theta$ near 60\si{\degree} are topologically trivial. The flat
bands at CB edge are nearly 12-fold degenerate (Fig. 4(c)) (two sets of 6-fold
degenerate bands with $\sim 4$meV separation), and originate from the
doubly-degenerate bands with antiparallel spins at the Q valley of UBZ.

To further clarify the spin-character of the bands, the bands V1 and V2 near the VB edge are shown along the path
$\textrm{K}_{\textrm{M}}-\textrm{M}_{\textrm{M}}-\Kp_{\textrm{M}}$ in
the MBZ in the inset in Fig. 4(a).  As can be seen, the two sets comprising the
V1 bands cross each other at the $\textrm{M}_{\textrm{M}}$ point of the MBZ and
have opposite ordering at the $\Kp_{\textrm{M}}$ point.  Fig. 4(d)
shows the variation  with $\theta$ of the energy gap (E$_W$) at the $\textrm{K}_{\textrm{M}}$ 
point between these two sets of bands (brown line), and of the minimum energy
gap between V1 and V2 (E$_{12}$) (purple line). As can be seen from the figure, 
E$_{12}$ becomes negative for $\theta\geq$54.9\si{\degree}, and  
E$_W$, which also denotes the bandwidth of V1 as both the maximum and minimum value
of V1 are at $\textrm{K}_{\textrm{M}}$, decreases with increasing MSL size. 
It is worth emphasizing that E$_W$ is not a spin-orbit splitting but arises due to band folding.    
This becomes evident if we look for the atomic spin-orbit partner of the V1 state at the
K$_{\textrm{M}}$ point by calculating the matrix elements S$_{ij}$  as discussed
before. We find large S$_{ij}$ for states $\sim$440 meV below the corresponding
states at the VB edge (Fig. 4(e)). 

\begin{figure}[ht]
\centering
\includegraphics[scale=0.32]{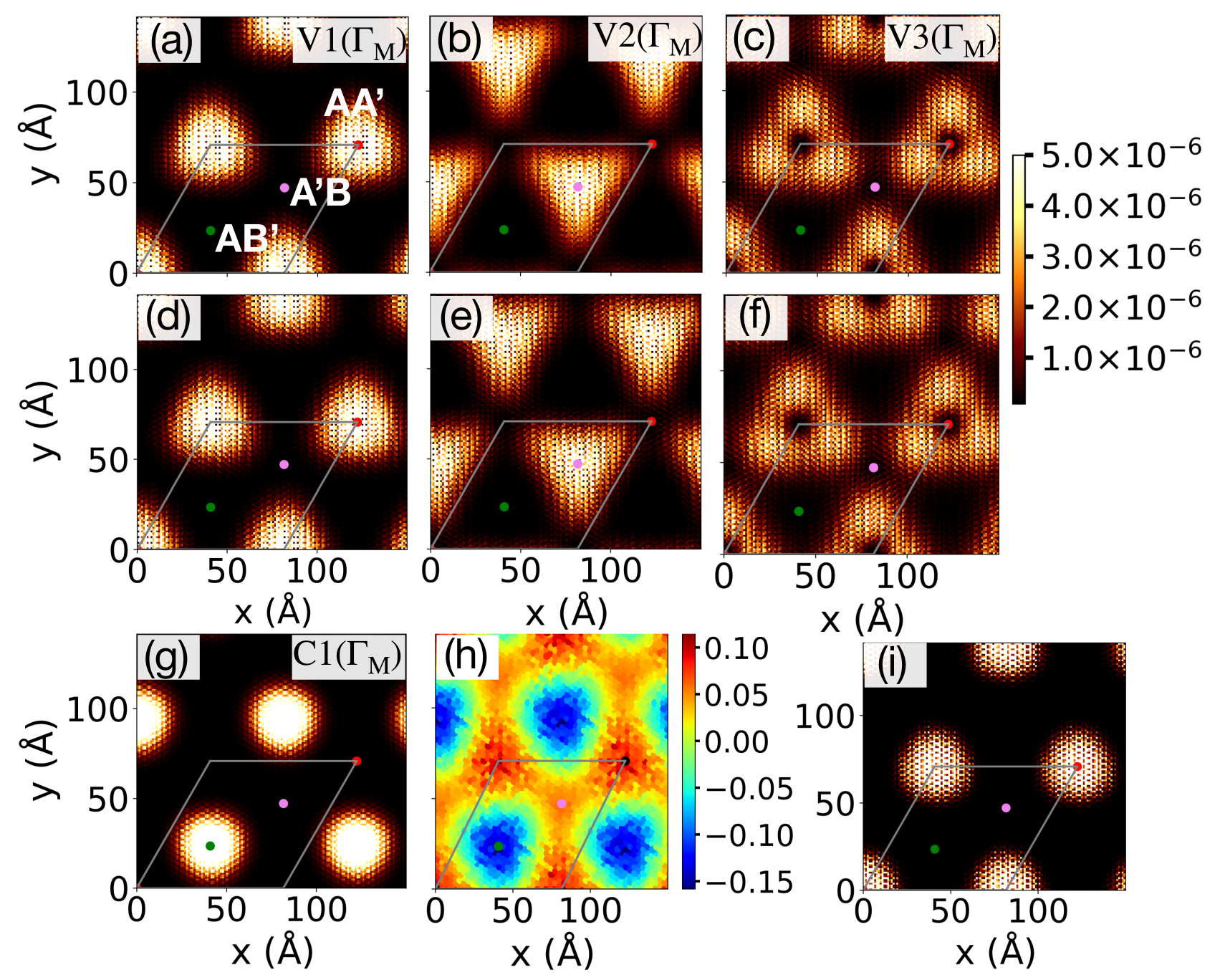}
\label{fig4}
\caption{(a), (b), and (c) ((d), (e), and (f)): Depiction of $|\psi_{\Gamma_{\textrm{M}}}(\textbf{r})|^2$ averaged
along the out-of-plane ($z$) direction for
V1, V2 and V3 (see Fig. 4)
bands from the relaxed-MSL bilayer (relaxed-MSL monolayer) for $\theta=$57.72\si{\degree} tWSe$_2$.
(g): Depiction of $|\psi_{\Gamma_{\textrm{M}}}(\textbf{r})|^2$ averaged
along the $z$ direction for the C1 band of the relaxed-MSL bilayer for $\theta=$57.72\si{\degree} tWSe$_2$.
(h): Moir\'{e} potential (in eV) for the relaxed-MSL monolayer.
(i): LDOS for bands arising from the $\Gamma$ point of the UBZ.}
\end{figure}

\textit{Wavefunction localization:}
The localization of the wavefunctions at the $\Gamma_{\textrm{M}}$ point of the MBZ
for the flat bands labelled V1, V2, V3 in Fig. 4(b)is depicted in Figs. 5(a)--(c),
and for the CB minimum (C1 of Fig. 4(c)), in Fig. 5(g). We find that V1 localizes
on the AA$^{\prime}$ stacking regions, whereas V2 and V3 localize on the A$^{\prime}$B and AA$^{\prime}$
stacking regions respectively. This is qualitatively different from what
happens in other twisted homobilayer TMDs \cite{naik1}. C1 has a very small
dispersion ($<$ 1 meV) and localizes strongly on the AB$^{\prime}$ stacking regions.

\textit{Comparison with relaxed monolayer:}
Some insight into the various features of the above band structure can be
obtained by noting that the local stacking at \textit{any} given point of the
tWSe$_2$ MSL with $\theta$ near 60\si{\degree} preserves the inversion
symmetry as seen in the high symmetry stackings: AA$^{\prime}$, AB$^{\prime}$ and A$^{\prime}$B 
(see \cite{sm_ref}). This inversion symmetry, together with
time-reversal symmetry, makes the effective interlayer interaction zero at the
K point of the UBZ \cite{continm}. It is also responsible for the trivial
topology of these bands.  In view of this, and the fact that the flat 
bands near the VB edge arise from the K point of the UBZ, we have studied the electronic
structure of the relaxed-MSL monolayer.  The results are shown as pink dashed
lines in Fig. 4(b) (VB edge) and Fig. 4(c) (CB edge). Clearly,
the flat bands near the VB edge originating from the bilayer MSL 
are remarkably similar to those from the relaxed MSL monolayer, with close agreement with respect to
the dispersion of the bands, as well as the separation between them.
This implies that the flat bands near the VB
edge originate primarily from the in-plane strains and not from the interlayer
hybridization. The first three wavefunctions at the VB edge of the relaxed MSL
monolayer are depicted in Figs. 5(d)--(f) and they look the same as those of the
MSL bilayer. On the other hand, the flat bands at the CB edge are derived from
the Q valley of the UBZ. Although the interlayer hybridization is nonzero at
the Q valley, the bands near the CB edge of bilayer MSL and of the relaxed MSL
monolayer also agree very well. Furthermore, the CB edge wavefunction at the
$\Gamma_{\textrm{M}}$ point of the relaxed monolayer localizes at the same stacking regions as
that of the full MSL.

In order to further understand the wavefunction localization, we have
computed the moir\'{e} potential ($V_{M}$) of the relaxed MSL monolayer (Fig.
5(h)).  $V_{M}$ has maxima at the AA$^{\prime}$ and A$^{\prime}$B regions and minima at the AB$^{\prime}$
regions, which is consistent with the holes localizing at AA$^{\prime}$ and A$^{\prime}$B and the
electrons localizing at AB$^{\prime}$. Hence, in the case of tWSe$_2$ with $\theta$ close
to 60\si{\degree} a relaxed MSL monolayer is sufficient to account for the electronic 
properties of the MSL if one is interested in the first few bands at the band edges.

\textit{Flat bands arising from the $\Gamma$ point of the UBZ:}
The monolayer band states near the $\Gamma$ point of the UBZ also give rise to
flat bands in the MBZ. The flat band at 129 meV {\em below the VB maximum} (marked as
V$_{\Gamma}$ in Fig. 4(b)) is the topmost such band. This band has a bandwidth $<$ 1
meV, and its wavefunction localizes strongly on the AA$^{\prime}$ regions, similar to the
findings in twisted MoS$_2$ \cite{naik1,naik2}. The localization is consistent with the ordering of
the VBs at the $\Gamma$ point in the UBZ of the different high symmetry stacking bilayers
\cite{sm_ref}. This V$_{\Gamma}$ band is two-fold
degenerate and has charge density in the interlayer region, unlike the states
arising from the K point of the UBZ. The LDOS corresponding to the state at the $\Gamma_{\textrm{M}}$
point of this band is shown in Fig. 5(i). STM studies on 57.5\si{\degree} tWSe$_2$ find $\Gamma$
derived bands localized on AA$^{\prime}$ stacking \cite{wse2_expt1} regions. Our findings on the
nature and localization of the $\Gamma$ derived flat band are thus in excellent agreement 
with the STM results.

\textit{Concluding comments:} In this letter we have presented 
an extensive study of the formation of flat bands in
tWSe$_2$ for $\theta$ near 0\si{\degree} as well as 60\si{\degree}, including 
the localization characteristics of the wavefunctions of these bands, using DFT. Our
study includes the strong spin-orbit coupling present in this system, as well
as the atomic rearrangements arising from the relaxation of the rigidly rotated
bilayer system. We find topologically nontrivial bands for tWSe$_2$ near
0\si{\degree} where interlayer hybridization together with the moir\'{e}
potential determines the electronic structure. For $\theta$ near
60\si{\degree}, the moir\'{e} potential of the relaxed MSL monolayer alone can account
for the electronic structure and the localization characteristics of the first
few flat bands at the band edges.  Furthermore, we have identified another set of
flat bands arising from the $\Gamma$ point of the UBZ. Our findings are in
excellent agreement with recent STM experiments. We believe that our study provides generic 
insights into the nature of flat bands in twisted TMD bilayers with strong
spin-orbit interactions.  The K-derived flat bands at the VB edge should prove
especially valuable for investigating the spin-valley physics in twisted
homobilayer TMD systems.

\begin{acknowledgments}
We thank the Supercomputer Education and Research Centre (SERC) at IISc
for providing the computational resources.
S. K. and M. J. acknowledge discussions with Johannes Lischner and Valerio Vitale.
M. J. gratefully acknowledges
financial support through grant no. DST/NSM/R\&D\_HPC\_Applications/2021/23 from the
National Supercomputing Mission of the Department of Science and Technology, India.
H. R. K. gratefully acknowledges the Science and Engineering Research Board of the Department of Science and Technology, India for financial support under grant No. SB/DF/005/2017.
\end{acknowledgments}
\end{document}


\author{Sudipta Kundu}
\author{Mit H. Naik}
\altaffiliation[Present address: ]{Department of Physics, University of California at Berkeley, California 94720, USA and Materials Sciences Division, Lawrence Berkeley National Laboratory, Berkeley, California 94720, USA.}
\author{H. R. Krishnamurthy}
\author{Manish Jain}
\affiliation{Center for Condensed Matter Theory, Department of Physics, Indian Institute of Science, Bangalore 560012, India}



\title
{Supplemental material for moir\'{e} induced topology and flat bands in twisted bilayer WSe$_2$: a first-principles study} 

\maketitle

\section{Computational Details and Methodology:}
We use the Twister code \cite{naik1} to generate the twisted bilayer WSe$_2$ (tWSe$_2$). As
the twist angle ($\theta$) approaches 0\si{\degree} or 60\si{\degree}, the number of atoms
in the moir\'{e} supercell becomes very large (Table 1).  Relaxing the twisted structure is
important to obtain the correct electronic structure and localization. Owing to
the large number of atoms, relaxation of these large systems using standard
density functional theory (DFT) calculations is challenging.  Here, we employ a
multiscale approach to relax our systems and calculate the electronic structure
of the relaxed systems, as discussed below.

\begin{table}[b]
\caption{Number of atoms in the moir\'{e} supercell and moir\'{e} length with varying $\theta$}
\begin{tabular}{|c|c|c|}
\hline
Twist angle($\theta$)       & No. of atoms & Length (\AA{}) \\ \hline
7.3\si{\degree}, 52.7\si{\degree}   & 366          & 25.39  \\ \hline
6.0\si{\degree}, 54.0\si{\degree}   & 546          & 31.01  \\ \hline
5.1\si{\degree}, 54.9\si{\degree}   & 762          & 36.63  \\ \hline
4.4\si{\degree}, 55.6\si{\degree}   & 1014          & 42.26  \\ \hline
3.9\si{\degree}, 56.1\si{\degree}   & 1302         & 47.89  \\ \hline
3.5\si{\degree}, 56.5\si{\degree}   & 1626         & 53.51  \\ \hline
3.14\si{\degree}, 56.86\si{\degree}   & 1986         & 59.14  \\ \hline
2.88\si{\degree}, 57.12\si{\degree} & 2382         & 64.77  \\ \hline
2.65\si{\degree}, 57.35\si{\degree} & 2814         & 74.80  \\ \hline
2.28\si{\degree}, 57.72\si{\degree}  & 3786         & 81.66  \\ \hline
1.89\si{\degree}, 58.11\si{\degree}  & 5514         & 98.55  \\ \hline
\end{tabular}
\end{table}

The unit cell lattice parameter of WSe$_2$ is 3.25 \AA{} in all our
calculations. We relax our twisted structures using classical force fields,
employing LAMMPS package \cite{lammps}. The intralayer
interaction is described by the Stillinger-Weber force field \cite{sw} and the
interlayer interaction by the Kolmogorov-Crespi forcefield
\cite{kc,kc_naik} which is fitted to van der Waals (vdW) corrected DFT calculations.
The forces are minimized using the conjugate gradient method till the force on
each atom becomes less than 10$^{-8}$ eV/\AA{}.

With the relaxed moir\'{e} structure, we perform DFT calculations to obtain the
electronic structure.  Our DFT calculations are performed using the
atomic-orbital based code: SIESTA \cite{siesta}. We use the polarized-double-zeta
basis set to expand the wavefunctions. We use norm-conserving pseudopotentials
and the exchange-correlation is treated within the local density
approximation (LDA) \cite{pz}.  As vdW correction in DFT only affects the
relaxation of the structure, we do not use vdW correction further in our electronic
structure calculations.  10\AA{} vacuum is added to restrict interaction
between periodic images in the out-of-plane direction.  We use a
12$\times$12$\times$1 Monkhorst-Pack k-point sampling \cite{kpt} to converge
the ground state charge density for the bilayer unit cell. For the moir\'{e} cell, we
accordingly scale the k-grid, and the large moir\'{e} cells are only sampled at the
$\Gamma_{\textrm{M}}$ point of their moir\'{e} Brillouin zone (MBZ).  All DFT
calculations include spin-orbit interaction.

We have performed GW calculations on the untwisted bilayer systems corresponding to different
high symmetry stackings using the BerkeleyGW package
\cite{BGW} with a 12$\times$12$\times$1 Monkhorst-Pack k-grid.  The mean-field
wavefunctions for the GW calculations are generated using Quantum Espresso
\cite{qe}.  To calculate the interlayer separation for the unit cells, we used
the DF-C09 \cite{dfc09} van der waals functional and the exchange-correlation
was treated using LDA. The  dielectric (epsilon) matrix is expanded in a plane-wave
basis with energy up to 15 Ry and extended to finite frequencies using the
generalized plasmon pole model \cite{gpp}. The Coulomb interaction is truncated
to compute the dielectric matrix, and the self-energy \cite{cellslab}. We use 1000
bands (without spin-orbit coupling) to construct the epsilon matrix and 1000
spinors for calculating the self-energy.  We use the static remainder technique
\cite{deslippe} to accelerate the convergence of the self energy with respect
to empty bands. While the number of bands is insufficient to obtain the
absolute convergence of the band edges with respect to vacuum, it is sufficient
to calculate the relative position of the valence band (VB) maximum at the $\Gamma$
point and the K point in the unit cell Brillouin zone.  To check for the convergence of the epsilon matrix
with respect to the number of bands, we have done calculations for the AB stacking
with 5000 bands (without spin-orbit coupling) and 1000 spinors for the construction of
the epsilon matrix. Our calculations show that using 1000 bands for computing the epsilon matrix is
sufficient to converge the relative energy difference. Table 2. shows our
results for the different high symmetry stackings. In all stackings,
the VB maximum at the K point of the unit cell Brillouin zone (UBZ) is higher in energy than at the $\Gamma$ point of the same.  

Furthermore, we have optimized the AB stacking structure with the generalized gradient
approximation \cite{pbe} and the DF-C09 \cite{dfc09} van der Waals functional. Even
in this case, our GW calculations show that the VB maximum at the K point is higher than
at the $\Gamma$ point by 191 meV.

\begin{table}
\caption{The VB energies at the $\Gamma$ and K points are reported. The first
entry for each stacking is for the first VB, and the second one is for the next
VB. All energies are in eV. The structures are optimized with LDA (except for the last row), and the
interlayer interaction is described by vdW-DFC09. The epsilon
matrix for the GW calculations is constructed with 1000 bands, and the quasiparticle
energies are calculated using 1000 spinors for the first five rows. The sixth and seventh
rows are calculated with 5000 bands and 1000 spinors respectively in the epsilon calculations. The epsilon matrix of the last row is again calculated with 1000 bands. }
\begin{tabular}{|c||c|c|c||c|c|c|}
\hline
    & \multicolumn{3}{c||}{DFT}  & \multicolumn{3}{c|}{GW}  \\ \hline
    & E$_{\Gamma}$  &  E$_K$  &  $\Delta$(=E$_K$ - E$_{\Gamma}$) &  E$_{\Gamma}$  &  E$_K$  &  $\Delta$(=E$_K$ - E$_{\Gamma}$) \\ \hline
AA$^{\prime}$ (2-H)  & 1.96 & 2.06 & 0.10 & -0.12 & 0.0012 & 0.12 \\
  & 1.96 & 2.06 & 0.10 & -0.12 & -0.0005 & 0.12 \\ \hline
  AB$^{\prime}$ (BW-W) & 1.89 & 2.07 & 0.18 & -0.21 & 0.02 & 0.23 \\
    &  1.89 & 2.07 & 0.18 & -0.21 & 0.02 & 0.23 \\ \hline
  A$^{\prime}$B (BSe-Se) & 1.72 & 2.07 & 0.35 & -0.42 & 0.0052 & 0.42 \\
    & 1.72 & 2.07 & 0.35  &  -0.42 & 0.0042 & 0.42 \\ \hline
  AA (3-R) & 1.71 & 2.10 & 0.39 & -0.15 & 0.28 & 0.43 \\
    &  1.71 & 2.03 & 0.32   & -0.15 &  0.23 & 0.38 \\ \hline
    AB (BW-Se) & 1.96 & 2.10 & 0.14 & -0.13 & 0.06 & 0.19 \\
    & 1.96 & 2.03 & 0.07 & -0.13 & -0.0058 & 0.13 \\ \hline
    & \multicolumn{6}{c|}{5000 bands for constructing epsilon matrix} \\ \hline
    AB (BW-Se) & 1.96 & 2.10 & 0.14 & -0.20 & -0.0027 & 0.20 \\
     & 1.96 & 2.03 & 0.07 & -0.20 & -0.07 & 0.13 \\ \hline
      & \multicolumn{6}{c|}{1000 spinors for constructing epsilon matrix} \\ \hline
 AB (BW-Se) & 1.96 & 2.10 & 0.14 & 0.21 & 0.41 &  0.20 \\
   & 1.96 & 2.03 & 0.07 & 0.21 & 0.34 & 0.13 \\ \hline
   & \multicolumn{6}{c|}{The structure is optimized with GGA and vdW-DFC09. } \\
   & \multicolumn{6}{c|}{1000 bands are used to construct epsilon matrix.} \\ \hline
   AB (BW-Se) & 0.96  & 1.08 & 0.12 & 0.0006 & 0.19 & 0.19 \\
 & 0.96  & 1.02& 0.06 & 0.0006 & 0.12 & 0.12 \\ \hline

\end{tabular}
\end{table}

\begin{figure}[h]
\centering
\includegraphics[scale=0.45]{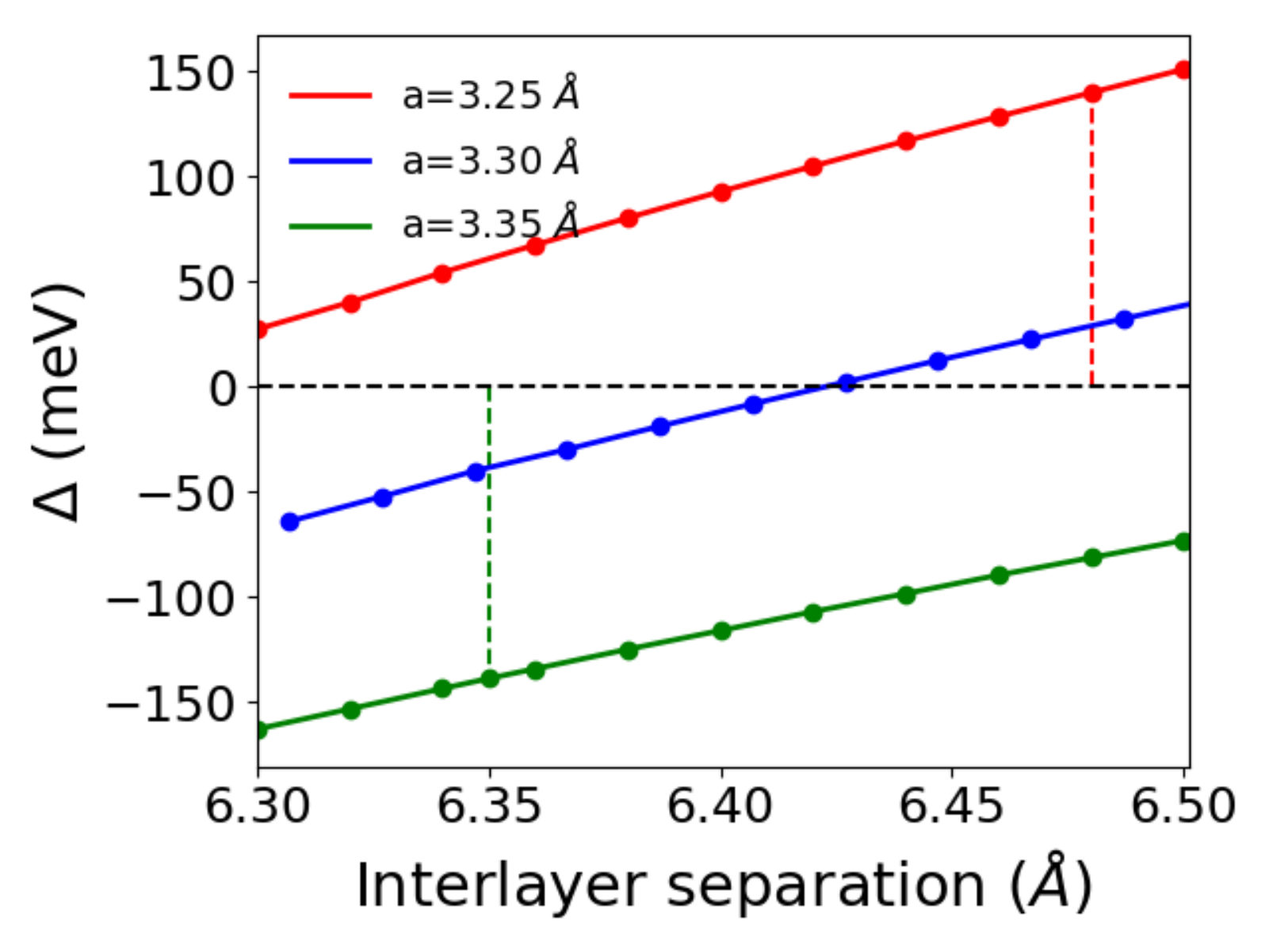}
\caption{Variation of $\Delta$ with interlayer separation for three different lattice constants (a). The vertical dashed lines
represent the equilibrium interlayer separation at a particular lattice constant.
The equilibrium interlayer separation for a = 3.30 \AA{} is 6.43 \AA{} at which $\Delta$ is 2.2 meV. Hence,
the blue dashed line is not visible.}
\end{figure}

It is to be noted here that the energy difference between the VB maximum at the K and
$\Gamma$ points ($\Delta\equiv$ E$_K$ - E$_{\Gamma}$) depends strongly on the
interlayer separation or the distance between the inner Se atoms.
Furthermore, the distance between W and Se also influences $\Delta$.  The
interlayer separation and $\Delta$ are very sensitive to strain. In Fig 1. we show the variation
of $\Delta$ with the interlayer separation for three different lattice constants (a). We find
that the $\Delta$ decreases with increasing lattice constant for a given value
of interlayer separation. $\Delta$ changes sign at a smaller value of interlayer separation with increasing lattice
constant. The changes in sign of $\Delta$ for a = 3.25 \AA{} and a = 3.35 \AA{}
occur outside the range shown in Fig. 1.  Furthermore, the W-Se distance and Se-Se
distance are also dependent on the functional used. A different functional, for
example Grimme's D2 functional, leads to smaller interlayer separation, which can lead to different
ordering of the VB edge at K and $\Gamma$.

\section{High Symmetry Stackings in Moir\'{e} Superlattice} 
Moir\'{e} superlattices (MSLs) in transition metal dichalcogenides have different sets of high symmetry stackings depending on
whether $\theta$ is close to 0\si{\degree} or 60\si{\degree}.  For $\theta$
close to 0\si{\degree}, the MSL has three high symmetry stackings: AA (W on top of W and Se on top
of Se), AB (W on top of Se) and BA (Se on top of W) 
(Figs. 2 (a--b)). The staggered stackings AB and BA are of lower energy than
the eclipsed stacking AA. In AA stacking, the Se atoms from both layers face
each other, and due to steric hindrance, the interlayer separation is
the maximum, at 7.1 \AA{}. In the moir\'{e} supercells, the interlayer separation is minimum (6.5
\AA{}) for the AB and BA stackings. The interlayer separation of any given stacking in WSe$_2$ is
larger than that of MoS$_2$ because of the denser electron cloud in Se compared
to S. In Figs. 3 (a--c), we show band structures of these high-symmetry
structures. Additionally, we show the band structure of the WSe$_2$ monolayer in
Fig.  3(b) (red dashed line). The VB maxima and conduction band (CB)
minima of all these bilayer structures are located at the K and Q points,
respectively. The VB maximum at the K point and the next band have parallel spins
(Fig. 3(a)), and there is a splitting of 30--50 meV between them depending on
the stacking. The split due to spin-orbit coupling is 445 meV which is also
present in the monolayer (Fig. 3(b)). Spin-orbit splitting is also present at the CB minimum
at the Q point. Additionally, there is interlayer splitting due to hybridization in
the bilayers, which is absent in the monolayer \cite{layerdep}. In this regard,
it is to be noted that the interlayer splitting of the VB maximum at the K point is very small as
the bands at the K point have very localized wavefunctions. On the other hand, the
CB minimum wavefunctions at the Q point and the VB maximum wavefunctions at the $\Gamma$ point are more
delocalized, leading to larger interlayer splittings at these points.

\begin{figure}[h]
\centering
\includegraphics[scale=0.45]{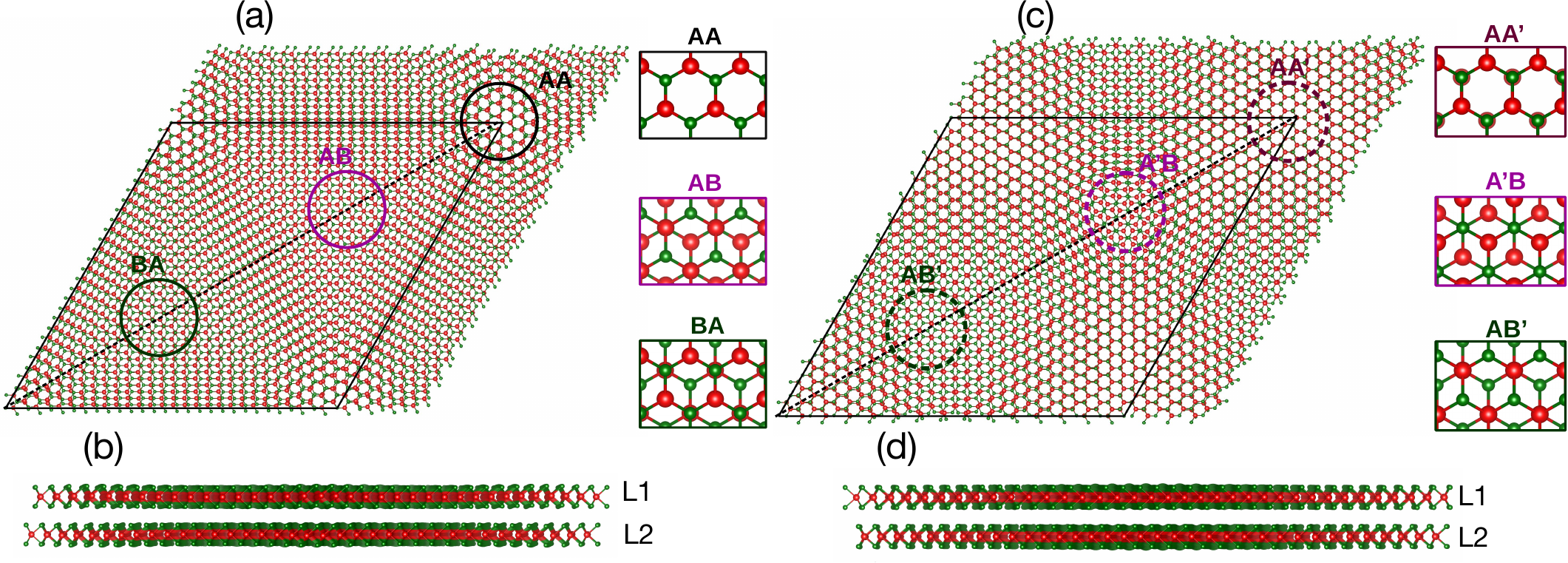}
\caption{Relaxed MSL of tWSe$_2$ with $\theta$ values of 2.28\si{\degree} (a) and 57.72\si{\degree} (c). The rectangular boxes show
different high symmetry stackings present in these structures. (b) and (d): Side views of (a) and (c) respectively. W and Se
atoms are shown in red and green colours respectively.}
\label{fig1}
\end{figure}

The other set of MSL strcutures with $\theta$ close to 60\si{\degree} contains AA$^{\prime}$ (W on
top of Se and Se on top of W), A$^{\prime}$B (Se on top of Se) and AB$^{\prime}$ (W on top of W) high symmetry stackings
(Figs. 2 (c--d)).  AA$^{\prime}$ (also known as 2H stacking) is the most stable structure
of bilayer WSe$_2$ and has an interlayer separation of 6.5 \AA{}. In A$^{\prime}$B stacking, Se atoms from both
layers face each other, leading to an increased interlayer separation of 7.1 \AA{} due to
electron-electron repulsion. A$^{\prime}$B stacking hence has the highest energy among these
three high symmetry stackings. Figs. 3 (d--f) show band structures of AA$^{\prime}$, AB$^{\prime}$ and A$^{\prime}$B,
respectively. Due to the presence of inversion symmetry in these high symmetry stackings, the VB maximum
are two-fold degenerate at the K point and have antiparallel spin (Fig. 3(d)).
The CB minima at the Q point are doubly degenerate.

\begin{figure}[h]
\centering
\includegraphics[scale=0.45]{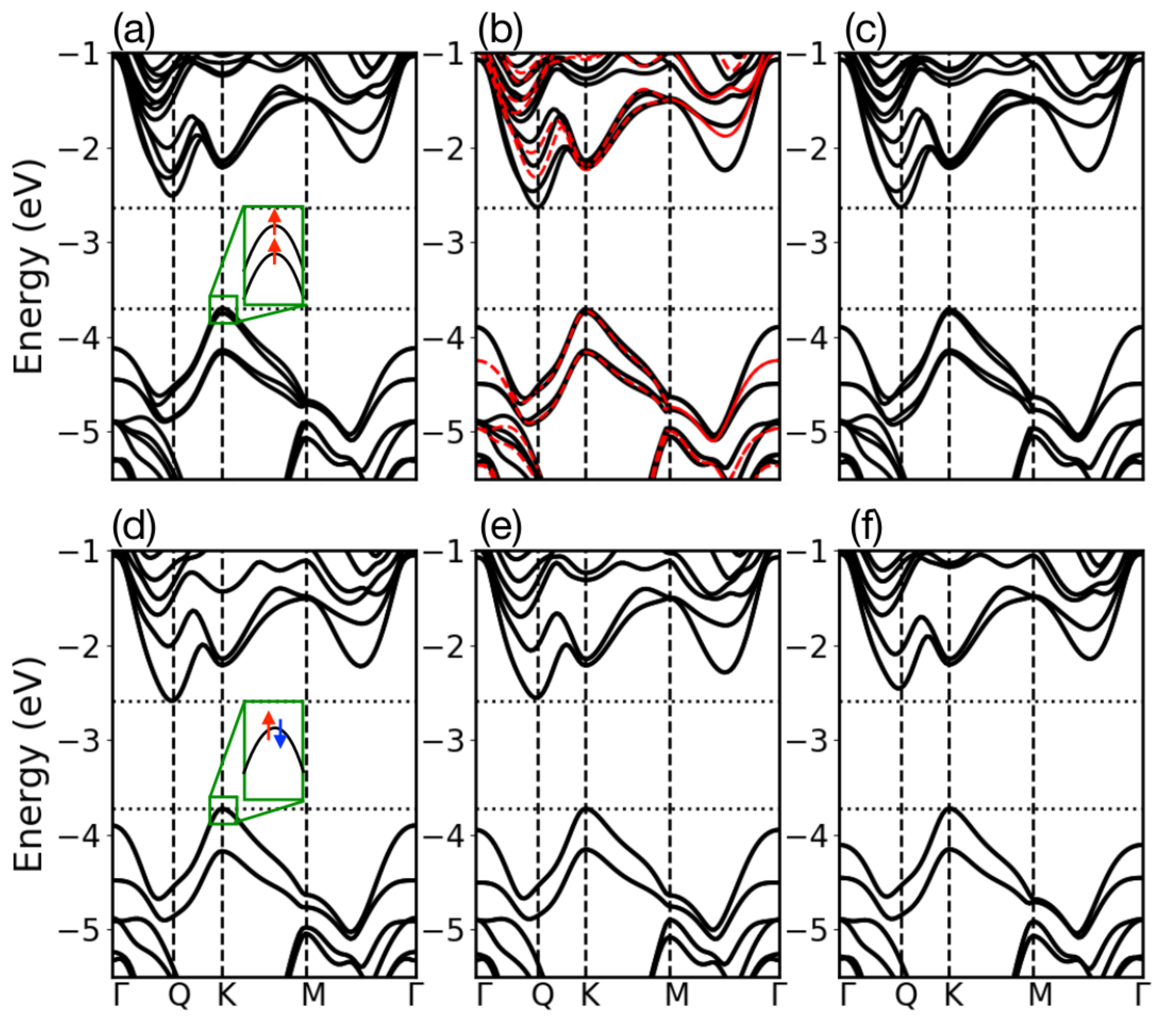}
\caption{Band structures of (a) AA, (b) AB, (c) BA, (d) AA$^{\prime}$, (e) AB$^{\prime}$ and (f) A$^{\prime}$B high symmetry stacked bilayers. The insets in (a) and (d) show the spin
configuration of the top two bands at the VB edge at the K point. The red dashed lines in (b) represent the band structure of the monolayer.}
\end{figure}

\section{Structural aspects}
In this section, we discuss the relaxed MSLs.  As the eclipsed stacking (AA) region has higher
energy compared to the staggered stacking regions (AB and BA, which are
degenerate), upon relaxation, the AB, BA regions expand equally while the AA
region shrinks, giving rise to a hexagonal pattern in MSL with
$\theta$ close to 0\si{\degree}. The relaxation effect becomes more evident as
$\theta$ approaches 0\si{\degree}. The relaxation pattern can be depicted by the
variation of the interlayer separation. We plot the distribution of the interlayer separation in Fig. 4(a).
The interlayer separation varies
between 6.5--7.1 \AA{}. AB and BA have the minimum interlayer separation. The relaxation process
also induces local strain and corrugation in the MSL. As a result, the
structure of the MSL is determined by the competition between strain-energy cost
and stacking-energy gain.  Figs. 4(b) and (c) show the strain distribution in
layer 1 (L1) and layer 2 (L2), respectively.  We plot the strain for every W
atom  along the direction of its six nearest W neighbours. In both layers, the
strains are zero at the high symmetry stackings and are localized in the domain walls. The strain
distribution in each layer is conjugate to the other {\em{i.e.}} at a given
point in the MSL along a certain direction, if the strain is compressive in L1, it
becomes tensile in L2.

\begin{figure}[ht]
\centering
\includegraphics[scale=0.4]{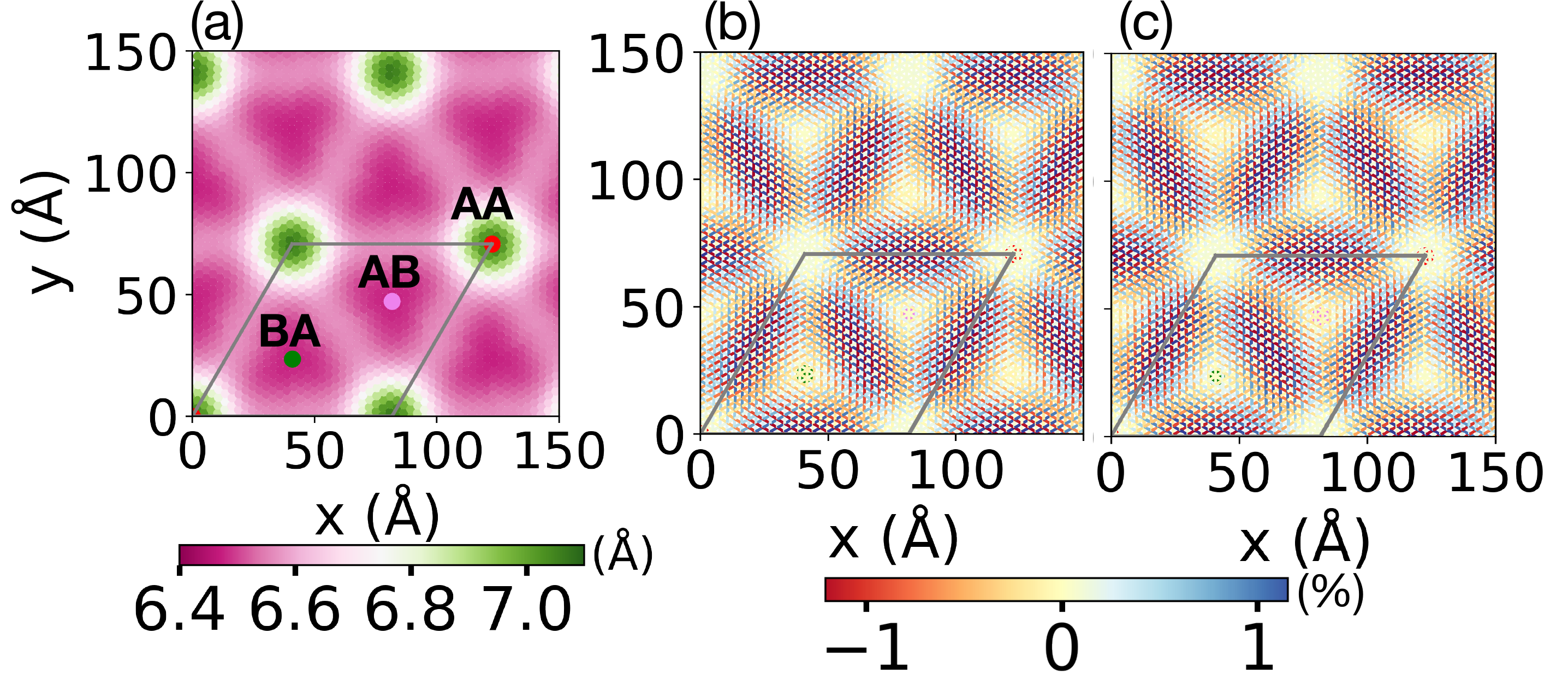}
\caption{(a): Distribution of the interlayer separation of tWSe$_2$ for $\theta$=2.28\si{\degree}. (b) and (c): Strain distributions in the two layers.}
\end{figure}
\begin{figure}[ht]
\centering
\includegraphics[scale=0.4]{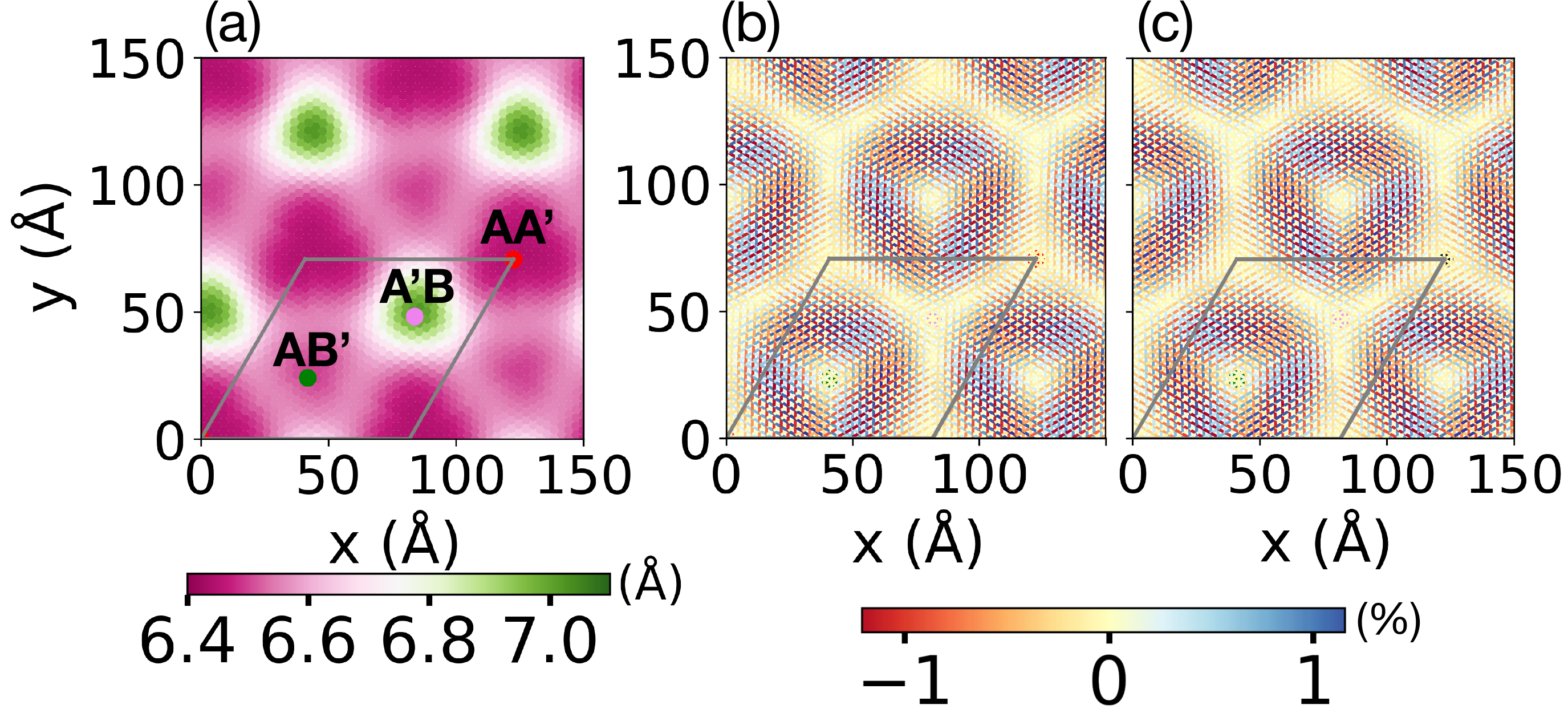}
\caption{(a): Distribution of the interlayer separation of tWSe$_2$ for $\theta$=57.72\si{\degree}. (b) and (c): Strain distributions in the two layers.}
\end{figure}
The MSL with $\theta$ close to 60\si{\degree} has AA$^{\prime}$, AB$^{\prime}$ and A$^{\prime}$B. The fact
that AA$^{\prime}$ has lower energy than AB$^{\prime}$ manifests in the reduced symmetry of the relaxed MSL,
and gives rise to Reuleaux triangles in the moir\'{e} patterns.
The highest energy stacking A$^{\prime}$B occupies the smallest area.
The interlayer separation varies between 6.5--7.1 \AA{}. The relaxation
pattern can be visualized by plotting the variation of interlayer separation in the MSL (Fig. 5(a)).
We also plot the distribution of strain induced by relaxing the MSL in Figs.
5(b) and (c).  We again find that the strains are localized on the domain walls
and are zero at the high symmetry stackings.  It is to be noted that lattice reconstruction, where
the lattice constant of the MSL changes, is not observed in the angle ranges we
have explored \cite{struc}.

\section{Calculation of Chern number}

The moir\'{e} Brillouin zones (MBZ) studied in our work are very small compared to the
UBZ. The flat bands at the VB edge in MBZ hence arise from monolayer
bands so close to the K and $\Kp$ points of UBZ that they can be associated
with up and down spins, respectively, with great accuracy.  (For example, in the
2.28\si{\degree} twisted WSe$_2$ (tWSe$_2$) the flat bands at the VB 
edge arise from K, $\Kp$ points and \textbf{k}-points within a circle around K, $\Kp$ point in UBZ
with the radius of $3.6\%$ of the reciprocal lattice vector of UBZ.) Due to the large spin-orbit
splitting at the K and $\Kp$ points, bands associated with the complementary spin will be $\sim$ 436 meV
below and contribute only to bands inside the VB continuum in the MBZ.
In particular, the flat bands at the VB edge in twisted WSe$_2$ with twist angle close
to 0\si{\degree} arise from the K point of the top layer and $\Kp$ point of the
bottom layer. As a result, the bands are doubly degenerate with antiparallel
spin, and we can differentiate between the flat bands arising from the K and $\Kp$
points by following the spin.

More specifically, we discretize the MBZ into a regular k-grid (in terms of the
reciprocal lattice vectors of the MBZ); for example, we sample the MBZ of
2.28\si{\degree} tWSe$_2$ with a 6$\times$6 k-grid. We compute the expectation value of the Pauli matrices at each grid point for the
first six bands at the VB edge. Fig. 6 depicts the k-grid in the hexagonal MBZ 
as well as the expectation value of
$\sigma_z$ ($\langle\sigma_z\rangle$) and $\sigma_x$ ($\langle\sigma_x\rangle$)
for the first two bands at the VB edge for 2.28\si{\degree} tWSe$_2$. As can be seen,
the value of $\langle\sigma_z\rangle$ is either close to $+$1 or $-$1  ($|\langle\sigma_z\rangle| > 0.9$)and
$\langle\sigma_x\rangle$ is very small ($|\langle\sigma_x\rangle| < 0.03$). This implies
that spin is essentially a good quantum number for the bands arising from the K/$\Kp$ points
of the UBZ. Hence we can identify the bands according to the $\langle\sigma_z\rangle$ value
in the whole MBZ and calculate the Chern number for the up spin
($\langle\sigma_z\rangle$ close to $+$1) and down spin ($\langle\sigma_z\rangle$
close to $-$1) separately. We find that the Chern numbers for the up and down
spin bands to be opposite, making the total Chern number of the system to be zero. It is
to be noted that purely energetically, the first band is spin-up at some $\mathrm{k}$-points in the MBZ
and spin-down at the rest of the $\mathrm{k}$-points (Fig. 6(a)). The other band
has the opposite spin to that of the first band (Fig. 6(b)).

\begin{figure}
\centering
\includegraphics[scale=0.4]{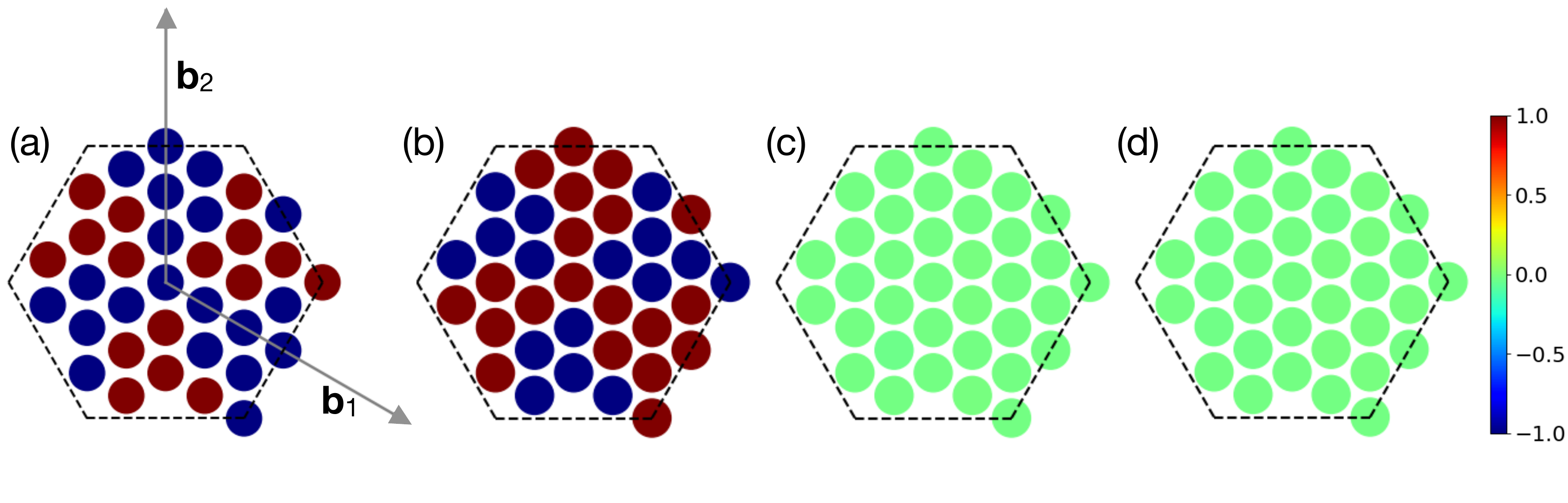}
\caption{(a) and (b): $\langle\sigma_z\rangle$ for the first two degenerate bands (V1 of Fig. 2(a) in main manuscript)
at VB edge 2.28\si{\degree} tWSe$_2$ in the hexagonal Brillouin zone. The reciprocal lattice vectors
are also shown in (a). (c) and (d): $\langle\sigma_x\rangle$ for the same bands.}
\end{figure}

To calculate the Chern numbers, we generate the Bloch wavefunctions for the various bands
on the above mentioned k-grid using density functional theory (DFT) calculations. 
As the MBZ is small, the 6$\times$6 sampling turns out to be sufficient for 2.28\si{\degree} tWSe$_2$ and divides
the MBZ into 36 plaquettes \cite{cherndft1,cherndft2}.
The Chern number for each spin (see discussion above) is then calculated following the equation:
\begin{equation}
C_{n,s} = \frac{1}{2\pi}\sum_{\mathbf{k}\in MBZ}\mathrm{Im} \left\{ \mathrm{ln}[\mathcal{U}^{n,s}_{\mathbf{k_1}\rightarrow\mathbf{k_2}} \mathcal{U}^{n,s}_{\mathbf{k_2}\rightarrow\mathbf{k_3}}\mathcal{U}^{n,s}_{\mathbf{k_3}\rightarrow\mathbf{k_4}} \mathcal{U}^{n,s}_{\mathbf{k_4}\rightarrow\mathbf{k_1}}] \right\}
\end{equation}
where $\mathcal{U}^{n,s}_{\mathbf{k_{\alpha}}\rightarrow\mathbf{k_{\beta}}}\equiv\frac{\langle\mathbf{u}_{n,s,\mathbf{k}_{\alpha}}\lvert\mathbf{u}_{n,s,\mathbf{k}_{\beta}}\rangle}{\lvert\langle\mathbf{u}_{n,s,\mathbf{k}_{\alpha}}\lvert\mathbf{u}_{n,s,\mathbf{k}_{\beta}}\rangle\lvert}$, $\alpha,\beta = 1,2,3,4$. $\mathbf{u}_{n,s,\mathbf{k}}$ is the periodic part of the Bloch wavefunction ($\psi_{n,s,\mathbf{k}}=\mathbf{u}_{n,s,\mathbf{k}}e^{i\mathbf{k}.\mathbf{r}}$) and $n,s$ refer to the
band number and spin of the band. $\mathbf{k_1},\mathbf{k_2},\mathbf{k_3},\mathbf{k_4}$ are the four vertices of each plaquette
ordered in an anti-clockwise fashion.
It is to be noted that we have employed different k-grids for different values of $\theta$ as the MBZ
becomes smaller with decreasing $\theta$; for example, the MBZ of 2.88\si{\degree} tWSe$_2$ is sampled with 
an 8$\times$8 k-grid. For 5.1\si{\degree} tWSe2, we checked the convergence of the C$_n$
by computing the C$_n$ with 12$\times$12 and 30$\times$30 k-grids and the C$_n$ are same for both.

\section{Comparison with experiments}
\begin{figure}
\centering
\includegraphics[scale=0.45]{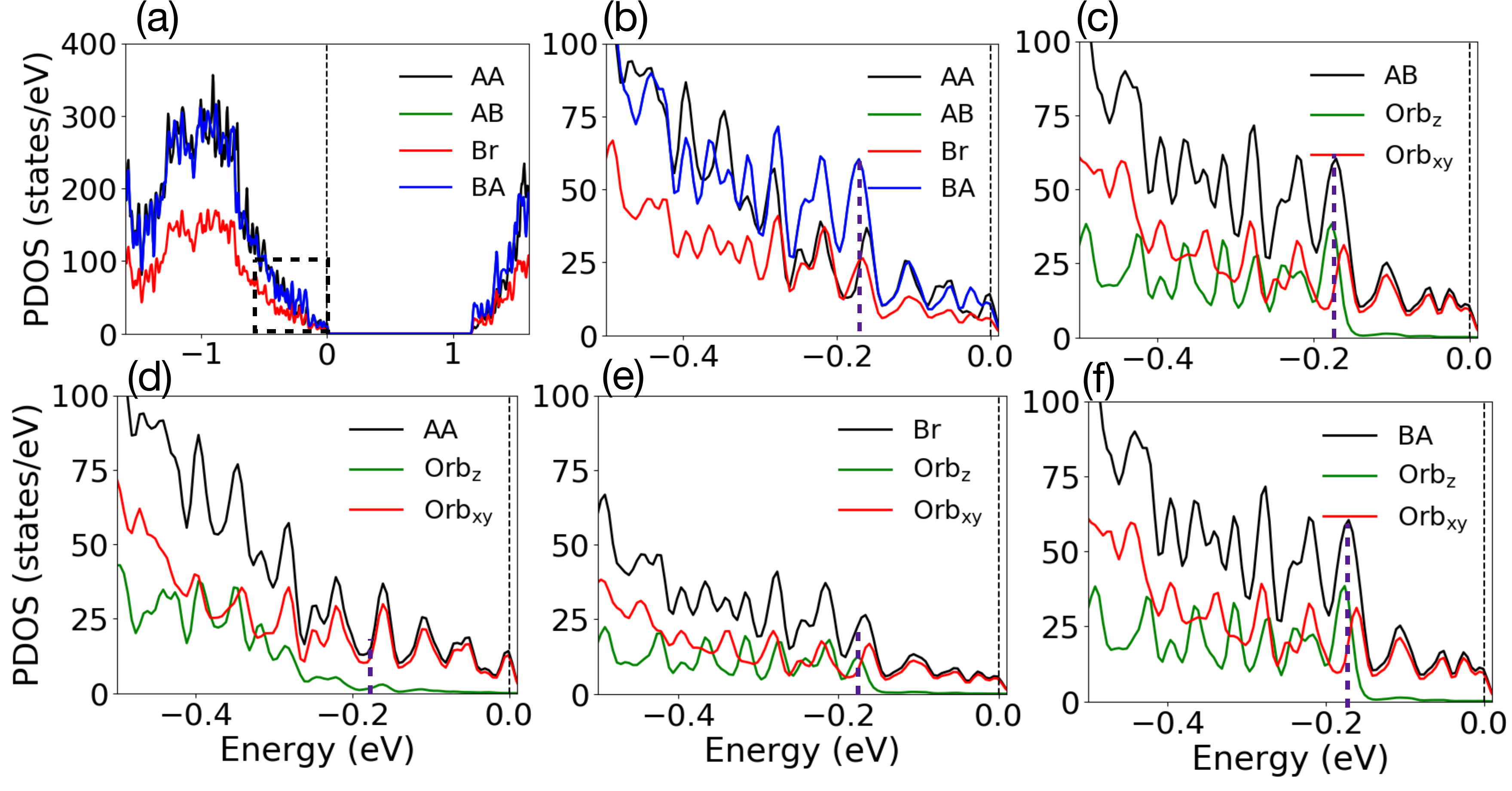}
\caption{(a): The PDOS contributions of AA, AB, Br and BA. (b): Enlarged view of PDOS
near the VB edge. The violet dashed
line marks the first peak due to $\Gamma$. (c)-(f): The contribution of Orb$_z$ (green line)
and Orb$_{xy}$ (red line) to the PDOS (black line) of
AB, AA, Br and BA respectively. Orb$_z$ consists of the d$_{z^2}$ orbital of W and the p$_z$ orbital of Se,
while Orb$_{xy}$ represents the d$_{x^2-y^2}$ and d$_{xy}$ orbitals of W atom and the p$_{x}$, p$_{y}$ orbitals of Se atom.}
\end{figure}

Scanning tunnelling measurements show flat bands coming from the $\Gamma$ point of
the UBZ for both kinds of MSL \cite{wse2_expt1}.  We have
shown the local density of states (LDOS) plots corresponding to those states in
the main manuscript.  To further compare with experimental findings, we calculate the
projected density of states (PDOS). We first consider a 2.88\si{\degree} tWSe$_2$ to compare with the STM on 3\si{\degree} tWSe$_2$.
We plot the PDOS of 2.88\si{\degree} tWSe$_2$ in Fig. 7(a). Fig. 7(b) shows the
zoomed in view of the PDOS near the VB edge. The first few states (up to
178 meV below the VB maximum of MSL) arise from the K point of the UBZ.  It is evident from
Fig. 7(b) that the first peak has a non-zero contribution from AA, AB, BA and Br (domain
walls between AB and BA). Hence, the LDOS calculated in the energy range of the first set
of doubly degenerate bands (V1 of Fig. 1(a) in main manuscript)
at the VB edge is delocalized in the whole MSL.  The violet dashed line in
Fig. 7(b) marks the first peak due to the band arising from the the $\Gamma$ point of
the UBZ. The states of this band are primarily localized in the AB, BA and Br regions.  The
wavefunction at the $\Gamma$ point of the UBZ consists of the $d_{z^2}$ orbital of the W atom
and $p_z$ orbitals of the Se atoms. We will refer to these orbitals as Orb$_z$. The
component along the out-of-plane direction of these orbitals makes the
interlayer interaction stronger, and hence larger interlayer splitting is found at
the $\Gamma$ point. On the other hand, the $d_{x^2-y^2}$ and $d_{xy}$ orbitals of the W atom and
the $p_x$, $p_y$ orbitals of the Se atoms (Orb$_{xy}$) contribute to the K point
wavefunctions of the UBZ. We plot the contributions of Orb$_z$ (green line) and
Orb$_{xy}$ (red line) to the PDOS (black line) of the AB and BA regions in Figs. 7(c) and (f) respectively.  The peak
marked by the violet dashed line has the main contribution from Orb$_z$, confirming
the fact that this peak arises from the $\Gamma$ point of the UBZ.  We also show the
contribution of Orb$_z$ and Orb$_{xy}$ to the PDOS of the AA and Br regions in Figs. 7(d)
and (e), respectively.

\begin{figure}
\centering
\includegraphics[scale=0.45]{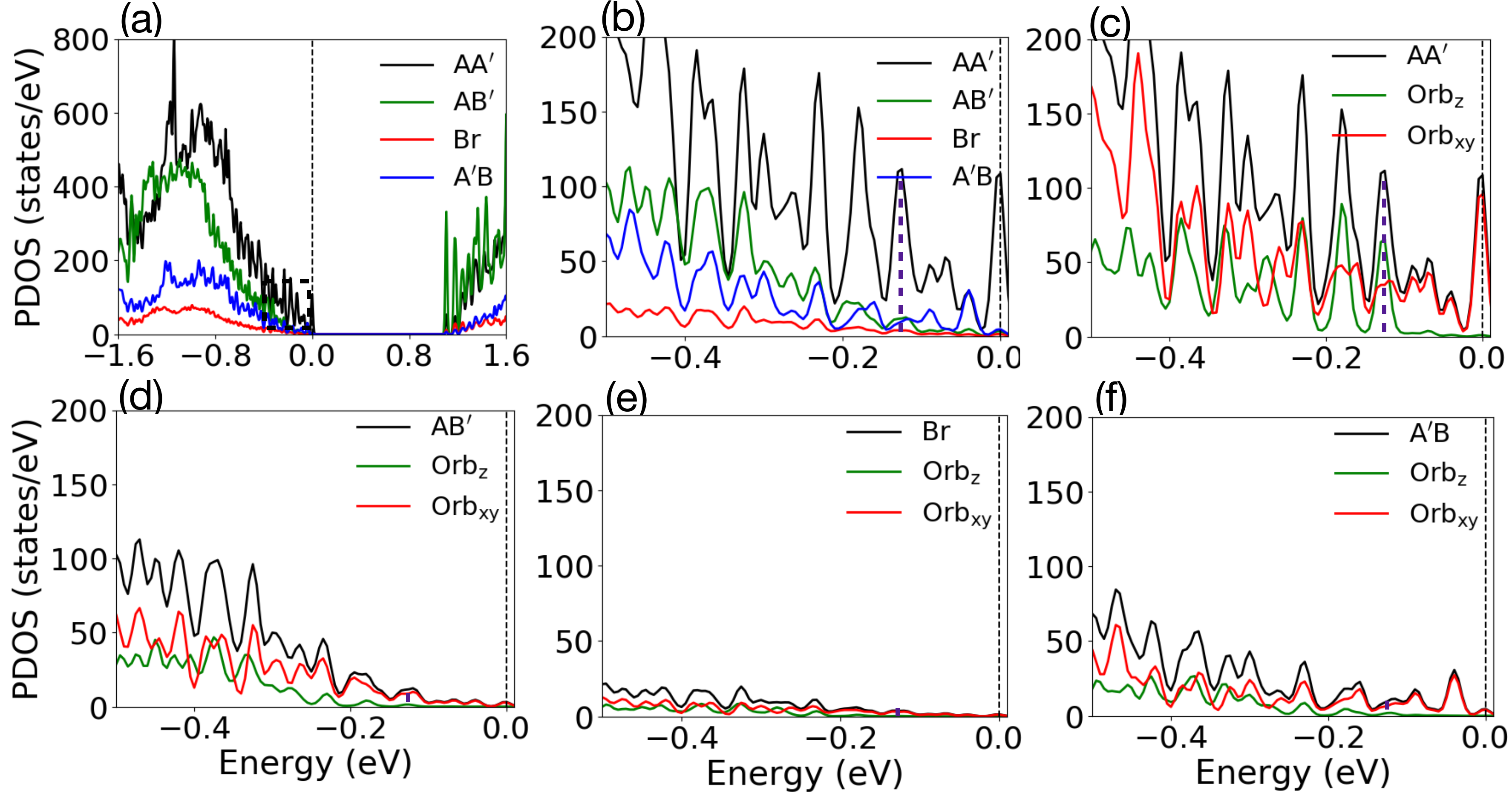}
\caption{(a)((b)): The PDOS (PDOS near the VB edge) due to AA$^{\prime}$, A$^{\prime}$B, Br and AB$^{\prime}$. The violet dashed line in (b) corresponds to the
peak arising due to the $\Gamma$ point. (c): The contribution of Orb$_z$ (green line) and Orb$_{xy}$ (red line) to the PDOS
(black line) of AA$^{\prime}$ to identify the $\Gamma$ and K
derived peaks. (d), (e) and (f): The orbital resolved
PDOS for A$^{\prime}$B, Br and AB$^{\prime}$.}
\end{figure}

Next, we consider the 57.72\si{\degree} tWSe$_2$ to compare with the STM on the
57.5\si{\degree} tWSe$_2$. We plot the PDOS in Fig. 8(a). Fig. 8(b) shows the
enlarged view of PDOS near the VB edge. The peak marked by the violet dashed
line in Fig. 8(b) arises from the $\Gamma$ point of the UBZ and has contributions from
AA$^{\prime}$ stacked regions, which implies that the $\Gamma$ derived flat band states localize
on AA$^{\prime}$. We plot the contributions of Orb$_z$ (green line) and Orb$_{xy}$ (red
line) to the PDOS (black line) of AA$^{\prime}$, AB$^{\prime}$, Br (bridge region between AA$^{\prime}$ and AB$^{\prime}$) and A$^{\prime}$B in
Figs. 8(c)--(f). The large contribution of Orb$_z$ to the peak marked by the violet
dashed line (Fig. 8(c)) shows the onset of the $\Gamma$ derived state. Furthermore,
this peak has a large contribution from the AA$^{\prime}$ stacking and negligible contributions from
other stackings.

\section{Moir\'{e} Hamiltonian}
The electronic structure at the VB edge of the relaxed MSL can be fit to a continuum model
\cite{continm} which can be used to go beyond the one-particle physics in these
systems.  The moir\'{e} Hamiltonian at the K valley for a particular spin state
can be written as:
\begin{equation}
H=
\begin{pmatrix}
\frac{-\hbar^2(\textbf{k}-{\kappa}_+)^2}{2m^*}+\Delta_{+}(\textbf{r}) & \Delta_{T}(\textbf{r})\\
\Delta^{\dagger}_{T}(\textbf{r}) & \frac{-\hbar^2(\textbf{k}-{\kappa}_-)^2}{2m^*}+\Delta_{-}(\textbf{r})
\end{pmatrix}
\end{equation}
Here, ${\kappa}_{\pm}$ are located at the MBZ corners and correspond to the K$_M$ and ${\Kp}_M$ points.
For $\theta$ close to 0\si{\degree},
the layer dependent moir\'{e} potential ($\Delta_{\pm}$) is given by:
\begin{equation}
\Delta_{\pm}(\textbf{r})=\sum_{j=1,3,5}2V\textrm{cos}(\textbf{G}_{j}.\textbf{r} \pm \psi)
\end{equation}
$V$ represents the potential amplitude and $\psi$ describes the spatial pattern.
$\textbf{G}_{j}$ is the $j^{th} $moir\'{e} reciprocal lattice vector in the first shell
obtained by an anti-clockwise rotation of the first vector \textbf{G}$_{1}$ by $(j-1)\pi/3$.
The interlayer tunneling ($\Delta_{T}$) is given by:
\begin{equation}
\Delta_{T}(\textbf{r})=w(1+e^{-i\textbf{G}_{2}.\textbf{r}}+e^{-i\textbf{G}_{3}.\textbf{r}})
\end{equation}
where $w$ is the interlayer tunneling strength.

\begin{figure}
\centering
\includegraphics[scale=0.45]{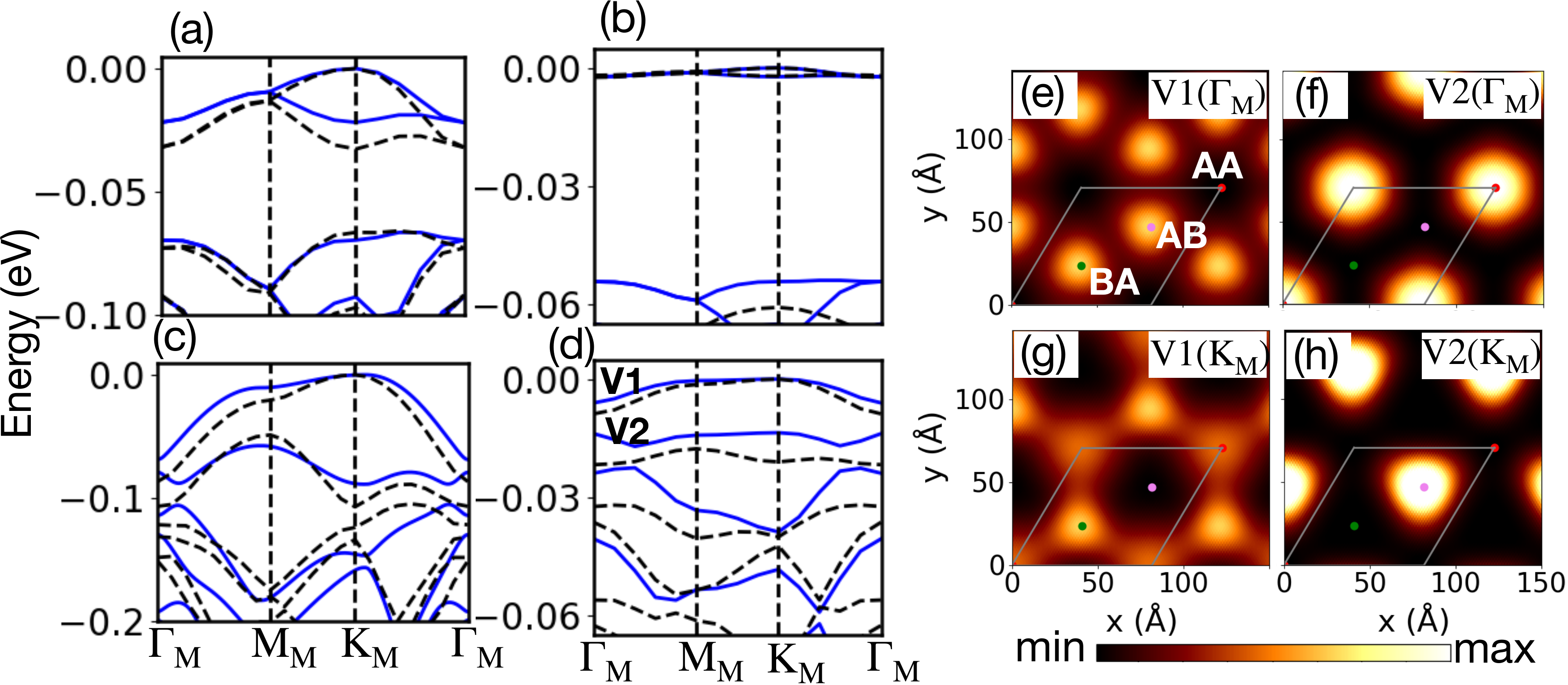}
\caption{(a) and (b): Band structures of 3.5\si{\degree} and 2.28\si{\degree} tWSe$_2$ with no interlayer tunneling
({\em{i.e.}} of the relaxed MSL monolayer). The blue solid and black dashed lines show the model band structure and the DFT result respectively.
(c) and (d): Band structures of 5.1\si{\degree} and 2.28\si{\degree} MSLs. The nearly-doubly-degenerate DFT bands have been averaged. 
(e) and (f) ((g) and (h)): Localization of wavefunctions of V1 and V2 (marked in Fig. (d)) at the $\Gamma_M$ (K$_M$) point.}
\end{figure}

First, we fit the model with no interlayer tunneling to the DFT results of
the relaxed MSL monolayer.  Figs. 9(a) and (b) (blue lines) show the band
structures obtained by solving the model with $w$ = 0 for 3.5\si{\degree} and
2.28\si{\degree}. For $V$ = 21 meV and $\psi$ = 94\si{\degree}, the model
band structures are in good agreement with DFT results. Fitting the model band
structures including the interlayer tunneling, we find $w$ = 30 meV. The band
structures with ($V$,$\psi$,$w$) = (21 meV, 94\si{\degree}, 30 meV) for
5.1\si{\degree} and 2.28\si{\degree} are shown in Figs. 9(c) and (d).  The Chern
numbers for the first two bands are consistent with the DFT results. We depict the
localization of the
wavefunctions of the first two spin-up bands of the 2.28\si{\degree} MSL (V1 and V2 in Fig.
9(d)) at the $\Gamma_{\textrm{M}}$ and K$_{\textrm{M}}$ points in Figs. 9(e)--(h) and the results
are in good agreement with the localization of the DFT wavefunctions (Fig. 3 in the main
manuscript). Hence, the model is able to describe the first two bands. A
limitation of the model would seem to be that neither the Chern number nor the localization characteristics
of the third band agrees with the DFT results. It is to be noted that F. Wu
et al. \cite{continm} gives a different set of parameters obtained from the DFT calculation on bilayer
unit cells. Although the parameters are very different, the Chern numbers and the localization of the
first two bands are consistent with our DFT and model calculations.
Widely different parameters can describe the same
band structure because of the fact that the bands are mostly determined by
kinetic energy. The intralayer and interlayer potentials act as perturbations
to split and modulate the bands.

\begin{figure}
\centering
\includegraphics[scale=0.45]{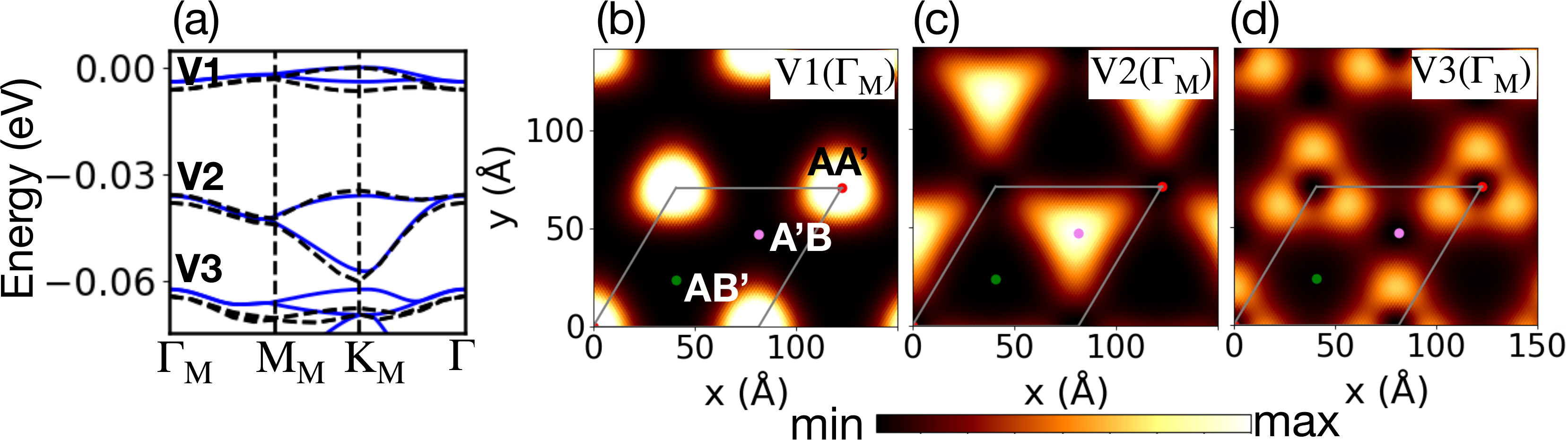}
\caption{(a): Band structures of 57.72\si{\degree} tWSe$_2$. Blue solid line shows the model band structure. (b)--(d):
Localization of V1, V2 and V3 at $\Gamma_M$.}
\end{figure} 

For $\theta$ close to 60\si{\degree}, the intralayer potential becomes:
\begin{equation}
\Delta(\textbf{r})=\sum_{j=1,3,5}2V\textrm{cos}(\textbf{G}_{j}.\textbf{r} + \psi)
\end{equation}
and the effective interlayer tunneling is zero. Hence, fitting the model with
DFT results for the 57.72\si{\degree} tWSe$_2$ (Fig. 10(a)), we find the parameters
to be ($V$,$\psi$,$w$) = (26 meV, 46\si{\degree}, 0). Furthermore, we show the localization of the
wavefunctions in Figs. 10(b)--(d) and they are in excellent agreement with the DFT
results.
%